\documentclass{article}

\usepackage[margin=2.5cm]{geometry}
\usepackage{latexsym}
\usepackage{appendix}
\usepackage{lineno}
\modulolinenumbers[5]
\usepackage{amsmath}
\usepackage{amssymb}
\usepackage{graphicx,caption}
\usepackage{ifpdf}
\usepackage[T1]{fontenc}
\usepackage{times}
\usepackage{float}
\usepackage{amsthm}
\usepackage{algcompatible}
\usepackage{algorithm}
\usepackage[noend]{algpseudocode}
\usepackage{xcolor}
\usepackage{xparse}
\usepackage{booktabs}
\usepackage{subfig}
\usepackage{url}
\usepackage{hyperref}

\newcommand{\argmax}{\operatornamewithlimits{argmax}}
\newcommand{\argmin}{\operatornamewithlimits{argmin}}

\algnewcommand\algorithmicinput{\textbf{Input:}}
\algnewcommand\INPUT{\item[\algorithmicinput]}
\algnewcommand\algorithmicoutput{\textbf{Output:}}
\algnewcommand\OUTPUT{\item[\algorithmicoutput]}

\newcommand{\norm}[1]{\left\lVert#1\right\rVert}

\title{Three-Dimensional Alignment of Density Maps in Cryo-Electron Microscopy}
\author{Yael Harpaz\protect\footnote{Department of Applied Mathematics, School of Mathematical Sciences, Tel-Aviv University, Tel-Aviv ,Israel. \url{yaelharpaz1@mail.tau.ac.il}} \and Yoel Shkolnisky\footnote{Department of Applied Mathematics, School of Mathematical Sciences, Tel-Aviv University, Tel-Aviv ,Israel. \url{yoelsh@tauex.tau.ac.il} }}

\begin{document}

\maketitle

\begin{abstract}
	A common task in cryo-electron microscopy (cryo-EM) data processing is to compare three-dimensional density maps of macromolecules. In this paper, we propose an algorithm for aligning three-dimensional density maps that exploits common lines between projection images of the maps. The algorithm is fully automatic and handles rotations, reflections (handedness), and translations between the maps. In addition, the algorithm is applicable to any type of molecular symmetry without requiring any information regarding the symmetry of the maps. 
	We evaluate our alignment algorithm on publicly available density maps, demonstrating its accuracy and efficiency. The algorithm is available at \url{https://github.com/ShkolniskyLab/emalign}.
\end{abstract}

\section{Introduction}
Single particle cryo-electron microscopy (cryo-EM) is a method to determine the three-dimensional structure of biological macromolecules from their two-dimensional projection images acquired by an electron microscope~\cite{Singer2012}. 
In this method, a sample of identical copies of the investigated molecule is quickly frozen in a thin layer of ice, where each copy is frozen at an unknown random orientation. The frozen sample is imaged by an electron microscope, resulting in two-dimensional images, where each image is a tomographic projection of one of the randomly oriented copies in the ice layer. The goal of single particle cryo-EM is to determine the three-dimensional structure of the molecule from the acquired two-dimensional images.
A~common task in cryo-EM data processing is to compare two density maps of the same molecule. This is required, for example, for estimating the resolution of the maps, evaluating their Fourier shell correlation curve~\cite{van2005fourier}, or to analyze their different conformations. All these tasks require to first align two density maps, that is, to orient them in the same way in a common coordinate system. Due to the nature of the cryo-EM imaging process, the two density maps may differ not only in their three-dimensional orientation (that is, their ``rotation''), but may also have different handedness (namely, reflected relative to each other), and may be centered differently with respect to a common coordinate system.

In this paper, we propose an algorithm for aligning two density maps, which is fully automatic and can handle rotations, translations, and reflections between the maps. The algorithm requires as an input only the two density maps. In particular, it does not assume knowledge of any other information such as the symmetry of the maps.

Formally, let $\phi_1 : \mathbb{R}^3 \rightarrow \mathbb{R}$ and $\phi_2 : \mathbb{R}^3 \rightarrow \mathbb{R}$ be two volumes such that
\begin{equation}\label{eq:AlignmentDef}
    \phi_2(r) = \phi_1(Or-t),  
\end{equation}
where $r=(x,y,z)^T \in \mathbb{R}^3$, ${O\in\ O(3)}$ and ${t=(\Delta x, \Delta y, \Delta z)^T}\in \mathbb{R}^3$ ($O(3)$ is the group of all orthogonal transformations of the three-dimensional space, namely, rotations and reflections). The alignment problem is to estimate~$O$ and~$t$ given~$\phi_1$ and~$\phi_2$. The matrix~$O$ is known as the orientation parameter, and the vector~$t$ as the translation parameter. In practice, we only get samples of $\phi_1$ and $\phi_2$, arranged as three-dimensional arrays of size $n\times n\times n$, where $n$ is the resolution of sampling. In cryo-EM, $\phi_1$ and $\phi_2$ represent two reconstructions of the same underlying molecule that we would like to compare (such as two half maps from a refinement process). In principle, it is possible to approximate the solution to the alignment problem using exhaustive search, by generating a set of candidate pairs $(O_i,t_i)$, where $O_i\in O(3)$ and $t_i\in \mathbb{R}^3$, and finding the pair which ``best aligns''~$\phi_1$ to~$\phi_2$ in some chosen metric. The purpose of the alignment algorithm presented in this paper is to estimate the optimal alignment parameters in a fast and accurate way. 

The paper is organized as follows. In Section~\ref{sec:previous}, we review existing alignment algorithms. In Section~\ref{sec:outline}, we give a high level simplified description of our algorithm. A detailed description is then given in Section~\ref{sec:Rotations}. This description relies of a method for aligning a single projection image against a volume, a procedure which is described in Section~\ref{sec:projectionAlignment}. In Section~\ref{sec:implementation}, we discuss implementation considerations of the algorithm, and analyze its complexity. An optional procedure for refining the estimated alignment parameters in described in Section~\ref{sec:optimization}. In Section~\ref{sec:results}, we demonstrate numerically the properties and performance of our algorithm. Finally, in Section~\ref{sec:conclusions}, we discuss the properties and advantages of our algorithm.
	
\section{Existing methods}\label{sec:previous}
There exist several methods for three-dimensional alignment of molecular volumes. The Chimera software~\cite{chimera} offers a semi-automatic alignment method which requires the user to approximately align the volumes manually, and then refines this alignment using an optimization procedure. This means that a sufficiently accurate initial approximation for the alignment is required. Achieving this initial approximate alignment manually naturally takes time and effort, yet it is crucial for the success of Chimera's alignment algorithm. The alignment procedure implemented by Chimera maximizes the correlation or overlap function between the two volumes by using a steepest descent optimization. The iterations of this optimization stop after reaching convergence or after 2,000 steps. 

Another alignment method is the projection based volume alignment algorithm (PBVA)~\cite{YU201393}. This method aligns a target volume to a reference volume by aligning multiple projections of the reference volume to the target volume whose orientation is unknown. The PBVA algorithm is based on finding two identical projections, a projection $P_{1}$ from the reference volume and a projection $P_{2}$ from the target volume as follows. The reference volume is projected at some known Euler angles, resulting in a projection $P_{1}$, and the matching projection $P_{2}$ is found by maximizing the cross-correlation function between $P_{1}$ and a set of projections representing the possible projections of the target volume. The cross-correlation function is of five parameters -- three Euler angles and two translation parameters in the plane of the projection~$P_{2}$~\cite{RADERMACHER1994121}. Finally, the rotation between the volumes is estimated from the relation between the Euler angles corresponding to the projections~$P_{1}$ and~$P_{2}$. After estimating the rotation between the volumes, the translation between them is found using projection images from the target volume. The translation between the volumes is estimated by least-squares regression using the two translation parameters of each projection from the target volume, where a minimum of two projections is required for calculating the three-dimensional translation vector. Using multiple projection images to estimate the translation between the volumes makes the alignment more robust. 

The Xmipp software package~\cite{Xmipp} also offers a three-dimensional alignment algorithm. It is based on expanding the two volumes using spherical harmonics followed by computing the cross-correlation function between the two spherical harmonics expansions representing the volumes~\cite{CHEN2013235}. 
The process of expanding a volume into spherical harmonics is called the Spherical Fourier Transform (SFT) of the volume, where like the FFT algorithm, there exists an efficient algorithm for calculating the SFT ~\cite{CHEN2013235}. 
The process of calculating the cross-correlation function between the two spherical harmonics expansions of the volumes and estimating the rotation between the two volumes is implemented by a fast rotational matching (FRM) algorithm~\cite{Kovacs:jn0106}.
After estimating the rotation between the two volumes, the translation between them is found by using the phase correlation algorithm~\cite{phase_corr}. 

Finally, the EMAN2 software package~\cite{EMAN2} offers two three-dimensional alignment algorithms. In the first algorithm (implemented by the program \texttt{e2align3d}, now mostly obsolete), the rotation between the volumes is estimated using an exhaustive search for the three Euler angles of the rotation. First, the algorithm generates a set of candidate Euler angles with large angular increments. Then, the algorithm iteratively decreases the angular increments in the set of candidates in order to refine the resolution of the angular search~\cite{EMAN_Align3D}. A much faster tree-based algorithm is implemented in the program \texttt{e2proc3d}. This method performs three-dimensional rotational and translational alignment using a hierarchical method with gradually decreasing downsampling in Fourier space. In Section~\ref{sec:results}, we compare our algorithm to this latter algorithm, as well as to the fast rotational matching algorithm implemented by Xmipp.

\section{Outline of the approach}\label{sec:outline}
We are given two volumes $\phi_{1}$ and $\phi_{2}$ satisfying~\eqref{eq:AlignmentDef}. For simplicity, we assume for now that the volumes have no symmetry, and are related by rotation only (no translation nor reflection). We generate a projection image from~$\phi_{2}$, denoted $P$, corresponding to an orientation given by a rotation matrix $R$. Since $\phi_{1}$ and $\phi_{2}$ are the same volume up to rotation, we can orient $P$ relative to $\phi_{1}$, that is, we can find the rotation $\tilde{R}$ such that projecting $\phi_{1}$ in the orientation determined by $\tilde{R}$ results in the image~$P$. As we show below, it holds that $\tilde{R} = O R$, where $O$ is the transformation from~\eqref{eq:AlignmentDef}. Since $R$ and $\tilde{R}$ are known, we can estimate $O$ as $O=\tilde{R} R^{T}$.

In practice, it may be that $\tilde{R}$ is not determined uniquely by $P$, as for example, a volume may have two very similar views even if it is not symmetric. Moreover, the volumes to align are discretized and sometimes noisy, which introduces inaccuracies into the estimation of $O$. Thus, to estimate $O$ more robustly, instead of using a single image~$P$, we generate from $\phi_{2}$ multiple images $P_{1},\ldots, P_{N}$ with orientations $R_{1},\ldots,R_{N}$, align each $P_{i}$ to $\phi_{1}$ as above, resulting in estimates for $O$ given by  $O_{i} = \tilde{R}_{i} R_{i}^{T}$, and then estimate $O$ from all $O_{i}$ simultaneously by solving 
\begin{equation*}
	O = \argmin_{R} \sum_{i=1}^N ||O_i - R||^2_F,
\end{equation*}
where $\norm{\cdot}_{F}$ is the Frobenius matrix norm. In Section~\ref{sec:Rotations}, we give an explicit solution for the latter optimization problem.

\begin{figure}
	\centering
	\includegraphics[width=0.5\textwidth]{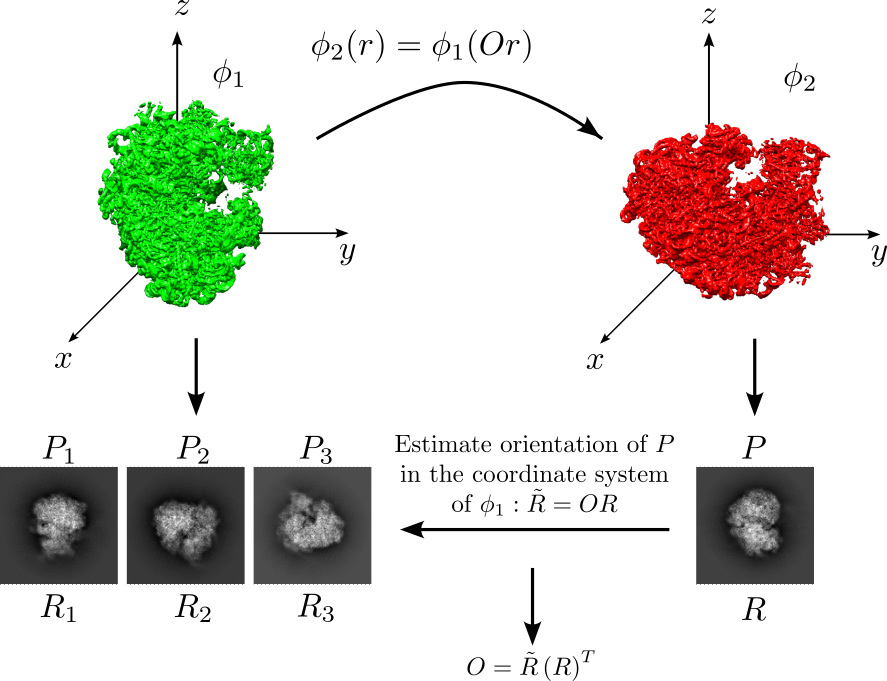}
	\caption{Outline of the algorithm}
	\label{fig:outline}
\end{figure}

The key of the above procedure is estimating the orientation of a projection image $P$ of $\phi_{2}$ in the coordinate system of~$\phi_{1}$. This is done by inspecting a large enough set of candidate rotations, and finding the rotation $\tilde{R}$ for which the induced common lines between $P$ (when assuming its orientation is $\tilde{R}$) and a set of projections generated from $\phi_{1}$ best agree. As inspecting each candidate rotation involves only one-dimensional operations (even if the input volumes are centered differently), it
is very fast and highly parallelizable. Thus, this somewhat brute-force approach is applicable to very large sets of candidate rotations (several thousands, for accurate alignment) and still results in a fast algorithm. We discuss below the complexity and advantages of this approach. We summarize the outline of our approach in Fig.~\ref{fig:outline}, and describe it in detail in Sections~\ref{sec:Rotations} and~\ref{sec:projectionAlignment}.

In the above approach, we assume that $O$ is a rotation. However, $\phi_{1}$ and $\phi_{2}$ may have a different handedness, and so $O$ may include a reflection. The above approach can obviously be used to resolve the handedness by aligning $\phi_{2}$ to $\phi_{1}$ and to a reflected copy of $\phi_{1}$, and determining whether a reflection is needed using some quality score of the alignment parameters (e.g., the correlation between the aligned volumes). However, as we show below, in our method, there is no need to actually align $\phi_{2}$ to a reflected copy of $\phi_{1}$, saving roughly half of the computations (those required to actually align $\phi_{2}$ to a reflected copy of $\phi_{1}$), as explained in Section~\ref{sec:Rotations}.

We next explain in detail the various steps of our algorithm, including handling translations, reflections, and symmetry in the volumes.

\section{Estimating the alignment parameters}\label{sec:Rotations}

Consider two volumes $\phi_1$ and $\phi_2$, where one volume is a rotated copy of the other (assuming for now  no reflection nor translation between the volumes), namely (see~\eqref{eq:AlignmentDef}) 
\begin{equation}\label{eq:OrotBetweenVols}
	\phi_2(r) = \phi_1(Or), \quad  r=(x,y,z)^T \in \mathbb{R}^3, 
\end{equation}
where $O$ is an unknown rotation matrix. Our goal is to find an estimate for $O$.

In case where $\phi_1$ and $\phi_{2}$ exhibit symmetry, the solution for $O$ is not unique. To be concrete, we denote by $SO(3)$ the group of all $3 \times 3$ rotation matrices. A group $G\subseteq SO(3)$ is a symmetry group of a volume $\phi$, if for all $g \in G$ it holds that 
\begin{equation}\label{eq:SymmetryDef}
	\quad \phi(r)=\phi(g r),\quad r=(x,y,z)^T \in \mathbb{R}^3.
\end{equation}
In other words, a symmetry group of a volume is a group of rotations under which the volume is invariant (see~\cite{van1999pointgroup} for more details). If we denote the symmetry group of $\phi_1$ by $G_1\subseteq SO(3)$ and define $r'=Or$, then, from \eqref{eq:OrotBetweenVols} and \eqref{eq:SymmetryDef}, we get for any symmetry element $g\in G_1$
\begin{equation}\label{eq:sym_est_for_O}
	\phi_2(r) = \phi_1(Or) = \phi_1(r') = \phi_1(gr') = \phi_1(gOr).
\end{equation}
Comparing the latter with~\eqref{eq:OrotBetweenVols}, we conclude that the solution for $O$ is not unique, and we thus replace the goal of finding $O$ by finding any $gO$ for some arbitrary element $g\in G_1$ of the symmetry group.

Note that we assume that $O$ is a rotation, namely that $\phi_1$ and $\phi_2$ are related by rotation without reflection. The case where $O$ is a reflection will be considered below.
Let $P$ be a projection image generated from $\phi_2$ using a rotation~$R$, that is
\begin{equation} \label{eq:projImage}
	P(x,y)=\int\limits_{-\infty}^\infty \phi(R r) dz= \int\limits_{-\infty}^\infty \phi(xR^{(1)}+yR^{(2)}+zR^{(3)})dz,
\end{equation}
where $R^{(1)}$, $R^{(2)}$, $R^{(3)}$ are the columns of the matrix $R$ and $r=(x,y,z)^T$. 
From \eqref{eq:OrotBetweenVols}, we have that
\begin{equation}
	\phi_2(R r) = \phi_1(O R r).
\end{equation}
Thus, using \eqref{eq:projImage}, we have
\begin{equation}\label{eq:O_relation_Pi}
	P(x,y) = \int\limits_{-\infty}^{\infty} \phi_2(R r)dz = \int\limits_{-\infty}^{\infty} \phi_1(OR r)dz.
\end{equation}
Equation \eqref{eq:O_relation_Pi} implies that if $P$ has orientation $R$ with respect to~$\phi_2$, then it has orientation $OR$ with respect to~$\phi_1$.  In Section~\ref{sec:projectionAlignment}, we describe how to estimate $O R$ given $P$ and~$\phi_{1}$, namely, how to estimate a rotation $\tilde{R}$ that satisfies $\tilde{R} = O R$. If the volume $\phi_{1}$ is symmetric with symmetry group $G_{1}$, then (as shown above) the rotation $O R$ is equivalent to the  rotation $g O R$ for any $g\in G_{1}$, and moreover, the two rotations cannot be distinguished. Thus, we conclude that 
\begin{equation*}
	\tilde{R} = g O R
\end{equation*}
for some unknown $g\in G_1$.
Using the latter equation, we can estimate $O$ as
\begin{equation}\label{eq:Xapprox}
	O = g^T \tilde{R} R^T.
\end{equation}
Note that in the latter equation $R$ is known, $\tilde{R}$ can estimated using the algorithm in Section~\ref{sec:projectionAlignment} below, and $g$ can be arbitrary. Thus, \eqref{eq:Xapprox}~provides a means for estimating~$O$. 

However, to estimate $O$ more robustly, we use multiple projections generated from~$\phi_{2}$. Let $R_{1},\ldots,R_{N}$ be  random rotations, and let $P_{1},\ldots,P_{N}$ be the corresponding projections generated from~$\phi_{2}$ according to~\eqref{eq:projImage}. Using the procedure described above, we estimate for each $P_{i}$ a rotation $\tilde{R}_{i}$ that satisfies $\tilde{R}_{i} = g_{i} O R_{i}$ for some unknown $g_{i} \in G_{1}$. Thus, as in~\eqref{eq:Xapprox}, we can estimate $O$ using any $i \in \left \{ 1,\ldots,N \right \}$ by
\begin{equation}\label{eq:O approx}
	O = g_{i}^{T} \tilde{R}_{i} R_{i}^{T}.
\end{equation}
Contrary to~\eqref{eq:Xapprox}, if we want the right hand side of~\eqref{eq:O approx} to result in the same $O$ for all $i=1,\ldots,N$, then~$g_{i}$ cannot be arbitrary. In order to estimate $O$, we therefore need to find~$g_i$, $i =  1,\dots,N$, and combine all estimates for~$O$ given in~\eqref{eq:O approx} into a single estimate.  

To that end, define 
\begin{equation}\label{eq:X_iNoReflect}
	X_i=R_i\tilde{R}_{i}^{T}, \quad  i = 1,\dots,N,
\end{equation} 
and look at the matrix $H$ of size $3N \times 3N$ whose $(i,j)$ block of size $3\times 3$ is given by (see~\eqref{eq:O approx} and~\eqref{eq:X_iNoReflect})
\begin{equation}\label{Xi^TXj_NoReflect}
	\begin{aligned}
		H_{ij} &= {X_i}^TX_j = \tilde{R}_i R_i^TR_j\tilde{R}_{j}^T \\
		&= \tilde{R}_i(g_i^T\tilde{R}_i)^T O O^T(g_j^T\tilde{R}_j)\tilde{R}_j^T 
		= {g_i}g_j^T.
	\end{aligned}
\end{equation}
By a direct calculation, we get that the matrix of size $3N \times 3$
\begin{equation}\label{eq:tilde g}
	\tilde{g} = \begin{pmatrix}
		g_1W\\
		\vdots \\
		g_NW\\
	\end{pmatrix},
\end{equation}
where $W$ is an arbitrary $3 \times 3$ orthogonal matrix (i.e., $W W^{T} = W^{T} W = I$) satisfies 
\begin{equation}\label{eq:eigenspace of H}
	H \tilde{g} = N \tilde{g}.
\end{equation} 
Equation~\eqref{Xi^TXj_NoReflect} also shows that the matrix~$H$ is of rank~3, which together with~\eqref{eq:eigenspace of H} implies that~$\tilde{g}$ can be calculated by arranging the three leading eigenvectors $v_{1}$, $v_{2}$, $v_{3}$ of $H$ in a matrix
\begin{equation}\label{eq:V_eigvec}
	V=\begin{pmatrix}
		\vline & \vline &\vline\\
		v_1 & v_2 & v_3\\
		\vline & \vline &\vline\\ 
	\end{pmatrix}_{3N\times 3} =\begin{pmatrix}
		V_1\\
		\vdots \\
		V_{N}
	\end{pmatrix}, 
\end{equation}
whose $3\times 3$ blocks $V_{1},\ldots,V_{N}$ are $g_{1}W,\ldots,g_{N}W$, for some unknown arbitrary $W$ (see~\cite{Eigenvector-synchronization} for a detailed derivation). In practice, at this point, we replace each $g_{i}W$ by its closest orthogonal transformation, as described in~\cite{arun1987least}, to improve its accuracy in the presence of noise and discretization errors.

Next, in order to extract an estimate for $g_{1},\ldots,g_{N}$ from~\eqref{eq:V_eigvec} (that is, to eliminate $W$ from the estimates in~$\tilde{g}$ given by~\eqref{eq:tilde g}), we multiply each $g_{i}W$ by $\left ( g_{1} W \right)^{T}$, resulting in   
\begin{equation}\label{eq:g_est_1}
	g_{est}= \begin{pmatrix}
		g_{est_1} \\
		\vdots \\
		g_{est_{N}}
	\end{pmatrix} = \begin{pmatrix}
		g_1W W^Tg_1^T \\
		\vdots \\
		g_NW W^Tg_1^T \\
	\end{pmatrix} = \begin{pmatrix}
		g_1{g_1}^T \\
		\vdots \\
		g_{N}{g_1}^T
	\end{pmatrix}.
\end{equation}
Thus, each $g_{est_i}$ is a rotation, even if $W$ is not.
We define $O_i = g_{est_i}^{T} \tilde{R}_{i} R_{i}^{T}$, and using~\eqref{eq:O approx}, we get for $i=1,\dots,N$
\begin{equation}\label{eq:Oi}
	\begin{aligned}
	O_i = g_{est_i}^{T} \tilde{R}_{i} R_{i}^{T} =g_1g_i^T\tilde{R}_iR_i^T = g_1 O.
\end{aligned}
\end{equation}
Thus, we have $N$ estimates for $g_1O$. 
Equation \eqref{eq:sym_est_for_O} states that $\phi_2(r) = \phi_1(Or) = \phi_1(gOr)$ for any symmetry element $g\in G_1$. Therefore, estimating $g_1O$ is equivalent to estimating $O$. 
In order to estimate $g_1O$ simultaneously from all $O_i$, $i =  1,\dots,N$, we search for the rotation $O_{est}^{(1)}$ (the superscript will be explained shortly) that satisfies
\begin{equation}\label{eq:O_est_optimization}
	O_{est}^{(1)} = \argmin_{R} \sum_{i=1}^N ||O_i - R||^2_F.
\end{equation}
In other words, $O_{est}^{(1)}$ is the ``closest'' to all the estimated rotations~$O_i$ in the least squares sense. To solve~\eqref{eq:O_est_optimization}, let~$\tilde{O}$ be the $3 \times 3$ matrix
\begin{equation}\label{Q_1}
	\tilde{O} = \frac{1}{N} \sum_{i=1}^N O_i.
\end{equation}
In \cite{singer2011three}, it is proven that the solution to the optimization problem in \eqref{eq:O_est_optimization} is 
\begin{equation}\label{eq:best_optimal_O}
	O_{est}^{(1)} = \tilde{U}\tilde{V}^T,
\end{equation}
where $\tilde{O}=\tilde{U}\tilde{\Sigma} \tilde{V}^T$ is the singular value decomposition (SVD) of~$\tilde{O}$.
The algorithm for estimating $O_{est}^{(1)}$ given $\phi_1$ and $\phi_2$, as described above, is presented in Algorithm~\ref{al:case_1}.

\begin{algorithm}
	\caption{Estimating $O_{est}^{(1)}$}\label{al:case_1}
	\begin{algorithmic}[1]
		\INPUT Volumes $\phi_1, \phi_2$
		\State{Generate random rotations $\{R_i\}_{i=1}^N$}
		\State{Generate from $\phi_2$ projections $\{P_{i}\}_{i=1}^N$  corresponding to the rotations $\{R_i\}_{i=1}^N$ \Comment Eq.~\eqref{eq:projImage}}
		\State{Apply Algorithm~\ref{al:Align2D} to each $P_{i}$ and $\phi_1$. Denote the resulting rotations by $\{\tilde{R}_i\}_{i=1}^N$}
		\For{$i=1$ to $N$}
		\State{Calculate $X_i = R_i\tilde{R}_i^T$ \Comment Eq.~\eqref{eq:X_iNoReflect}}
		\EndFor    
		\State{Construct the $3N\times 3N$ matrix
			\begin{equation*}
			H=\begin{pmatrix}
				I_{3\times 3} & X_1^T X_2 & \cdots & X_1^T X_N\\
				X_2^T X_1 & I_{3\times 3} & \cdots & X_2^T X_N\\
				\vdots & \vdots & \ddots &\\
				X_N^T X_1 & X_N^T X_2 & \cdots & I_{3\times 3}
			\end{pmatrix}
		\end{equation*}}
		\State{Find the three leading eigenvectors $v_1,v_2,v_3$ of $H$
			\State Set $$V=\begin{pmatrix}
				\vline & \vline &\vline\\
				v_1 & v_2 & v_3\\
				\vline & \vline &\vline\\ 
			\end{pmatrix}_{3N\times 3} =\begin{pmatrix}
				V_1\\
				\vdots \\
				V_{N}
			\end{pmatrix}$$ \Comment Eq.~\eqref{eq:V_eigvec}}
		\State{Compute $$g_{est}=\begin{pmatrix}
				g_{est_1} \\
				\vdots \\
				g_{est_{N}}
			\end{pmatrix}=\begin{pmatrix}
				V_1 V_1^T\\
				\vdots \\
				V_N V_1^T\\
			\end{pmatrix}$$ \Comment Eq.~\eqref{eq:g_est_1}}
		\For{$i=1$ to $N$}
		\State{Calculate $O_i = g_{est_i}^T \tilde{R}_{i} R_{i}^{T}$ \Comment Eq.~\eqref{eq:Oi}}
		\EndFor
		\State{Calculate $\tilde{O} = \frac{1}{N} \sum_{i=1}^N O_i$ \Comment Eq.~\eqref{Q_1}}
		\State{Calculate $O_{est}^{(1)} = \tilde{U}\tilde{V}^T$, where $\tilde{O}=\tilde{U}\tilde{\Sigma} \tilde{V}^T$ is the SVD of~$\tilde{O}$  \Comment Eq.~\eqref{eq:best_optimal_O}}
		\OUTPUT $O_{est}^{(1)}$ \Comment Estimated rotation
	\end{algorithmic}
\end{algorithm}

To handle the case where $\phi_{1}$ and $\phi_{2}$ have a different handedness (namely, related by reflection), we can of~course apply Algorithm~\ref{al:case_1} to $\phi_{2}$ and a reflected copy of $\phi_{1}$. However, this would roughly double the runtime of the estimation process, as the most time consuming step in Algorithm~\ref{al:case_1} is step~3, whose complexity is $O(n^{3}\log n)$ operations for a volume of size $n \times n \times n$ voxels (see Section~\ref{sec:projectionAlignment}).

Alternatively, it is possible to augment the above algorithm to handle reflections without doubling its runtime. In the case where there is a reflection between $\phi_{1}$ and $\phi_{2}$, we need to replace the relation in~\eqref{eq:OrotBetweenVols} by the relation
\begin{equation}\label{eq:ReflectionBetweenVols}
\phi_{2}(r) = \phi_{1}(O J r), \quad J=\begin{pmatrix}
		1 & 0 & 0\\
		0 & 1 & 0\\
		0 & 0 & -1
	\end{pmatrix}.
\end{equation}
Note that $J$ in~\eqref{eq:ReflectionBetweenVols} is a reflection and that $O$ is a rotation. Repeating the above derivation starting from~\eqref{eq:ReflectionBetweenVols} shows that to estimate $O$ in this case, we can use the same $R_{i}$ used above and the same estimates $\tilde{R}_{i}$ obtained above (steps~1 and~3 of Algorithm~\ref{al:case_1}), but this time we get that $O_{i} = g_{i}^{T} \tilde{R}_{i} J R_{i}^{T} J$ (compare with~\eqref{eq:O approx}). Then, we set $X_{i} = J R_{i} J \tilde{R}_{i}^{T}$ (compare with~\eqref{eq:X_iNoReflect}) and proceed as above, resulting in an estimate $O_{est}^{(2)}$ (compare with~\eqref{eq:O_est_optimization}), which corresponds to the optimal alignment parameters if $\phi_{1}$ and $\phi_{2}$ have opposite handedness. Once we have the two estimates $O_{est}^{(1)}$ and $O_{est}^{(2)}$ for the alignment parameters between $\phi_{1}$ and $\phi_{2}$ (without and with reflection), 
we estimate the translation corresponding to each of $O_{est}^{(1)}$ and $O_{est}^{(2)}$ using phase correlation~\cite{phase_corr} (see Appendix~\ref{app:ShiftEst3D} for details). This results in two sets of alignment parameters (rotation+translation). We then apply both sets of parameters to $\phi_{2}$ to align it with~$\phi_{1}$, and pick the parameters for which $\phi_{2}$ after alignment has higher correlation with $\phi_{1}$. We denote the estimated parameters by $(O_{est}, t_{est})$.

\section{Projection alignment}\label{sec:projectionAlignment}
It remains to show how to implement step~3 of Algorithm~\ref{al:case_1}, that is, how to find the orientation of a projection $P$ of $\phi_{2}$ with respect to the coordinate system of $\phi_{1}$. Mathematically, we would like to solve the equation
\begin{equation}\label{eq:ProjectionAlignment_no_shift}
	P(x,y)= \int\limits_{-\infty}^\infty \phi_{1}(Rr) dz, \quad r=(x,y,z)^T\in \mathbb{R}^3
\end{equation}
for the unknown rotation $R$. A brute-force approach of testing many candidate rotations in search for the $R$ that (best) satisfies~\eqref{eq:ProjectionAlignment_no_shift} is prohibitively expensive, as it requires to compute a projection of $\phi_{1}$ for each candidate rotation (this is essentially projection matching). We therefore take a different approach, whose cost for inspecting each candidate rotation is much lower (in fact requires $O(n)$ operations to test each candidate rotation for a volume $\phi_{1}$ discretized into an array of size $n \times n \times n$).

The idea is to generate several projection images from~$\phi_{1}$, and then, for each candidate rotation, to check the agreement of the common lines between $P$ and the projections of~$\phi_{1}$, assuming the orientation of $P$ is given by the candidate rotation. We estimate the rotation corresponding to~$P$ as the candidate rotation that results in the best agreement. We next formalize this method, and then analyze its complexity.

We start by considering the case where there is no translation between~$P$ and~$\phi_{1}$, namely,~$P$ and~$\phi_{1}$ satisfy~\eqref{eq:ProjectionAlignment_no_shift}, and our goal is to estimate~$R$ given~$P$ and~$\phi_{1}$. We generate~$N$ projection images from $\phi_{1}$ ($N$ is typically small, see Section~\ref{sec:results}), denoted $P_{1}^{(a)},\ldots,P_{N}^{(a)}$, with rotations $R_{1}^{(a)},\ldots,R_{N}^{(a)}$ chosen uniformly at random (note that we deliberately reuse the notation $N$ used in Section~\ref{sec:Rotations}, as explained below) . We generate a set of candidate rotations $S$, over which we will search for the solution~$R$ of~\eqref{eq:ProjectionAlignment_no_shift}. The set $S$ consists of a large number of approximately equally spaced rotations. See Appendix~\ref{app:candidates} for a detailed description of the construction of~$S$.

We will assume for each candidate rotation $Q\in S$ that~$P$ was generated using the rotation~$Q$ (that is, we assume that~$R$ in~\eqref{eq:ProjectionAlignment_no_shift} is equal to $Q$), compute the mean correlation of the common lines between $P$ and $P_{1}^{(a)},\ldots,P_{N}^{(a)}$, and choose as an estimate for $R$ the rotation $Q$ for which the mean correlation is highest. Specifically, for each $Q\in S$ and $R_{i}^{(a)}$, $i=1,\ldots,N$, we compute the direction of the common line between $P$ and $P_{i}^{(a)}$, given by the angles~$\alpha_{i}$ in~$P$ and~$\beta_{i}$ in~$P_{i}^{(a)}$, as explained in Appendix~\ref{app:common lines}. The common line property~\cite{shkolnisky2012viewing} states that if $Q=R$ then 
\begin{equation*}
	\hat{P}(\xi \cos \alpha_{i},\xi \sin \alpha_{i}) = \hat{P}_{i}^{(a)} (\xi \cos \beta_{i}, \xi \sin \beta_{i}),\quad \xi \in \mathbb{R},
\end{equation*}
where $\hat{P}$ and $\hat{P}_{i}^{(a)}$ are the Fourier transforms of $P$ and $P_{i}^{(a)}$, respectively (see Appendix~\ref{app:common lines} for a review of common lines and their properties). We thus define
\begin{align*}
		f_{i}(Q,\xi) &= \hat{P}(\xi \cos \alpha_{i},\xi \sin \alpha_{i}),\\ 
		g_{i}(Q,\xi) &= \hat{P}_{i}^{(a)} (\xi \cos \beta_{i}, \xi \sin \beta_{i}),
\end{align*}
and the cost function
\begin{equation}\label{eq:cost no shift}
	\rho(Q) = \frac{1}{N} \Re \sum_{i=1}^{N} \frac{\int_{0}^{\infty} \bar{f}_{i}(Q,\xi) g(Q,\xi)\, d\xi}{\norm{f_{i}}_{L^{2}} \norm{g_{i}}_{L^{2}}},
\end{equation}
where $\bar{f}_{i}$ denotes the complex conjugate of $f_{i}$.
In other words, $\rho(Q)$ measures how well the common lines induced by $Q$ between $P$ and $P_{1}^{(a)},\ldots,P_{N}^{(a)}$ agree. We then set our estimate for~$R$ to be
\begin{equation*}
	R_{est} = \argmax_{Q\in S} \rho(Q).
\end{equation*}
We explore the appropriate value for $N$ in Section~\ref{sec:results}.

We now extend the above scheme to the case where $P$ is not centered with respect to $\phi_{1}$, namely, $P$ is given by
\begin{equation} \label{eq:ProjectionAlignment}
	P(x-\Delta x,y-\Delta y)= \int\limits_{-\infty}^\infty \phi_{1}(Rr) dz, \quad r=(x,y,z)^T\in \mathbb{R}^3,
\end{equation}
for an unknown rotation $R$ and an unknown translation $(\Delta x, \Delta y)$. The idea for estimating $R$ is the same as before, except that the calculation of the common lines should take into account the unknown translation, as we describe next.

We denote the unshifted version of $P$ by $\tilde{P}$, which is given by 
\begin{equation}\label{eq:tilde P}
	\tilde{P}(x,y)= \int\limits_{-\infty}^\infty \phi_{1}(Rr) dz, \quad r=(x,y,z)^T\in \mathbb{R}^3
\end{equation}
(this is exactly~\eqref{eq:ProjectionAlignment_no_shift}, but we repeat it to clearly set up the notation). Then,
\begin{equation*}
	P(x,y) = \tilde{P}(x+\Delta x, y+ \Delta y).
\end{equation*}
Taking the Fourier transform of both sides of the latter equation, we get that~\cite{Singer2012}
\begin{equation}\label{eq:Fourier shift property}
	\hat{P}(\omega_{x},\omega_{y}) = \hat{\tilde{P}}(\omega_{x},\omega_{y}) e^{\imath (\omega_{x} \Delta x + \omega_{y} \Delta y)}.
\end{equation}
Suppose that the common line between $\tilde{P}$ and $P_{i}^{(a)}$ is given by the angles~$\alpha_{i}$ in~$\tilde{P}$ and~$\beta_{i}$ in~$P_{i}^{(a)}$ (see Appendix~\ref{app:common lines}). By definition of the common line, it holds that 
\begin{equation*}
	\hat{\tilde{P}}(\xi\cos\alpha_{i}, \xi\sin\alpha_{i}) = \hat{P}_{i}^{(a)} (\xi\cos\beta_{i}, \xi\sin\beta_{i}).
\end{equation*}
Using~\eqref{eq:Fourier shift property}, we get that
\begin{equation*}
	\hat{P}(\xi\cos\alpha_{i}, \xi\sin\alpha_{i}) e^{-\imath \xi \Delta \xi} = \hat{P}_{i}^{(a)} (\xi \cos \beta_{i}, \xi \sin \beta_{i}),
\end{equation*}
where $\Delta \xi = \Delta x \cos \alpha_{i} + \Delta y \sin \alpha_{i}$ is the one-dimensional shift between the projections along their common line. We assume that this one-dimensional shift is bounded by some number $d$.

Thus, we need to modify our cost function~\eqref{eq:cost no shift} to take into account also the unknown (one-dimensional) phase $e^{-\imath \xi \Delta \xi}$. We therefore define (with a slight abuse of notation in reusing the previous notation for the cost function)
\begin{align*}
	f_{i}(Q,\Delta \xi, \xi) &= \hat{P}(\xi \cos \alpha_{i},\xi \sin \alpha_{i}) e^{-\imath \xi \Delta \xi},\\ 
	g_{i}(Q,\xi) &= \hat{P}_{i}^{(a)} (\xi \cos \beta_{i}, \xi \sin \beta_{i})
\end{align*}
and the cost function
\begin{equation}\label{eq:cost function}
	\rho(Q,\Delta \xi) = \frac{1}{N} \Re \sum_{i=1}^{N} \frac{\int_{0}^{\infty} \bar{f}_{i}(Q,\Delta \xi, \xi) g(Q,\xi)\, d\xi}{\norm{f_{i}}_{L^{2}} \norm{g_{i}}_{L^{2}}},
\end{equation}
and set our estimate for the solution $R$ of~\eqref{eq:ProjectionAlignment} to be
\begin{equation}\label{eq:max rho}
	R_{est} = \argmax_{\substack{Q\in S, \ \Delta \xi \in [-d,d]}} \rho(Q,\Delta \xi).
\end{equation}
The formula for the angles $\alpha_{i}$ and $\beta_{i}$ of the common line between $P$ and $P_{i}^{(a)}$ induced by the rotations $Q\in S$ and $R_{i}$ is given in Appendix~\ref{app:common lines}. Note that at this point we are only interested in $R_{est}$ and not in the translation $(\Delta x, \Delta y)$ in $P$, as the relative translation between $\phi_{1}$ and $\phi_{2}$ is efficiently determined using phase correlation (see~\cite{phase_corr} and Appendix~\ref{app:ShiftEst3D}) once we have determined their relative rotation. The algorithm for solving equation~\eqref{eq:ProjectionAlignment} is summarized in Algorithm~\ref{al:Align2D}.

\begin{algorithm}
	\caption{Projection alignment}\label{al:Align2D}
	\begin{algorithmic}[1]
		\INPUT Projection $P$ and volume $\phi_{1}$ satisfying~\eqref{eq:ProjectionAlignment} 
		\State{Generate random rotations $R_{1},\ldots,R_{N}$}
		\State{Generate from $\phi_{1}$ projections $P_{1}^{(a)},\ldots,P_{N}^{(a)}$ corresponding to the rotations
		$R_{1},\ldots,R_{N}$}	
		\State{Generate candidate rotations $S$}\Comment{Appendix~\ref{app:candidates}}
		\State Compute
		\begin{equation*}
			R_{est} = \argmax_{\substack{Q\in S, \ \Delta \xi \in [-d,d]}} \rho(Q,\Delta \xi).		
		\end{equation*}\Comment{Eqs.~\ref{eq:cost function} and~\eqref{eq:max rho}}	\OUTPUT $R_{est}$ \Comment {Estimated rotation}
	\end{algorithmic}
\end{algorithm}

As mentioned above, we use the same $N$ in Sections~\ref{sec:Rotations} and~\ref{sec:projectionAlignment}. While in principle, the number of projections generated from~$\phi_{2}$ in Section~\ref{sec:Rotations} can be different from the number of projections generated from~$\phi_{1}$ in Section~\ref{sec:projectionAlignment}, due to the symmetric role of~$\phi_{1}$ and~$\phi_{2}$ in the alignment problem, there is no reason to consider different values.

\section{Implementation and complexity analysis}\label{sec:implementation}
Algorithms~\ref{al:case_1} and~\ref{al:Align2D} are formulated in the continuous domain. Obviously, to implement them, we must explain how to apply them to volumes $\phi_{1}$ and $\phi_{2}$ given as three-dimensional arrays of size $n \times n \times n$. We now explain how to discretize each of the steps of Algorithms~\ref{al:case_1} and~\ref{al:Align2D}, and analyze their complexity. For simplicity, we use for the discrete quantities the same notation we have used for the continuous ones. 

The only step in Algorithm~\ref{al:case_1} that needs to be discretized is step~2. This step is accurately discretized based on the Fourier projection slice theorem~\eqref{proj-sliceThem} using a non-equally spaced fast Fourier transform~\cite{FIN,B20}, whose complexity is $O(n^{3} \log n)$ (for a fixed prescribed accuracy). The result of this step is a discrete projection image $P$ given as a two-dimensional array of size $n \times n$ pixels. The remaining steps of Algorithm~\ref{al:case_1} are already discrete, and since the value of $N$ is small compared to $n$, their complexity is negligible.

We next analyze Algorithm~\ref{al:Align2D}. The input to this algorithm is a projection image $P$ of size $n\times n$ pixels, and a volume $\phi_{1}$ of size $n \times n \times n$ voxels. The algorithm also uses the parameter~$N$, but since it is a small constant, we ignore it in our complexity analysis. Step~1 of Algorithm~\ref{al:Align2D} requires a constant number of operations. Step~2 is accurately implemented using a non-equally spaced fast Fourier transform~\cite{FIN,B20}, whose complexity is $O(n^{3} \log n)$ (for a fixed prescribed accuracy). Step~3 is independent of the input volume, and moreover, the set $S$ can be precomputed and stored. To implement step~4, we first discretize the interval of one-dimensional shifts $[-d,d]$ in fixed steps of $\Delta d$ pixels (say, 1~pixel). Specifically, we use the following shift candidates for the optimization in step~4
\begin{equation*}
	\Delta \xi \in \left \{ -d + k \Delta d \ | \ k=0,\ldots, \left \lfloor 2d/\Delta d \right \rfloor \right \}.
\end{equation*}
Then, for each $Q \in S$, we compute the angles~$\alpha_{i}$ and~$\beta_{i}$ (see Appendix~\ref{app:common lines}), and evaluate~\eqref{eq:cost function} for the pair $(Q,\Delta \xi)$ by replacing the integral with a sum. If we store the polar Fourier transforms of all involved projection images~$P$ and $P_{1}^{(a)},\ldots,P_{N}^{(a)}$ (computed using the non-equally spaced fast Fourier transform~\cite{FIN,B20}), each such evaluation amounts to accessing the rays in the polar Fourier transform corresponding to the angles $\alpha_{i}$ and $\beta_{i}$, namely $O(n)$ operations. Thus, the total number of operations required to implement step~4 of Algorithm~\ref{al:Align2D} is $\lvert S \rvert \times (\lfloor 2d/\Delta d \rfloor + 1) \times n$ ($\lvert S \rvert$ is the number of elements in the set $S$). Of course, all $\lvert S \rvert \times (\lfloor 2d/\Delta d \rfloor + 1)$ evaluations are independent, and can be computed in parallel.
Thus, the total complexity of Algorithm~\ref{al:Align2D} is $O(n^{3} \log n)$ operations for step~2 and $O(n)$ operations for testing each pair $(Q,\Delta \xi)$ in step~4. Therefore, since the optimization in step~4 is very fast, it is practical to test even a very large set of candidate rotations~$S$. 

Finally, we note that in practice, to further speed up the algorithm, we first downsample the input volumes to size $n_{ds}$, align the two downsampled volumes, and apply the estimated alignment parameters to the original volumes. We demonstrate in Section~\ref{sec:results} that this approach still results in a highly accurate alignment.

To understand the theoretical advantage of the above approach, we compare it to a brute force approach. In the brute force approach, we 1)~scan over a large set of rotations and three-dimensional translations, 2)~for each pair of a rotation and a translation, we transform one of the volumes according to this pair of parameters, and 3)~choose the pair for which the correlation between the volumes after the transformation is maximal. Testing each pair of candidate parameters requires $O(n^{3})$ operations (for rotating and translating one of the volumes, and for computing correlation), which amounts to a total of $O(n^{3} \times \lvert{S}\rvert \times (2d/\Delta d)^{3})$ operations. In other words, testing each candidate rotation and translation is way more expensive than in our proposed method. In our approach, the expensive operation of complexity $O(n^{3} \log n)$ needs to be executed only once per each pair of inputs~$P$ and~$\phi_{1}$. Moreover, in our approach, the search over shifts is one-dimensional as opposed to the three-dimensional search required in the brute-force approach.

\section{Parameters' refinement} \label{sec:optimization}
In this section, we describe an optional refinement procedure for improving the accuracy of the estimated parameters $O_{est}$ and~$t_{est}$ obtained using the algorithm of Section~\ref{sec:Rotations}. 

We define the vector $\Theta = (\psi,\vartheta,\varphi,\Delta_x,\Delta_y,\Delta_z)$ consisting of the~$6$ parameters required to describe the transformation between two volumes --  $3$ Euler angles ($\psi,\vartheta,\varphi$) describing their relative rotation, and~$3$ parameters $(\Delta x,\Delta y,\Delta z)$ describing their relative translation. We define the operator $T_{\Theta}(\phi)$, which applies the transformation parameters~$\Theta$ to the volume~$\phi$ (that is,~$T_{\Theta}$ first rotates the volume and then translates it, according to the parameters in~$\Theta$). Next, for given volumes~$\phi_{1}$ and~$\phi_{2}$, we denote their correlation by $\rho(\phi_{1},\phi_{2})$. We are reusing the notation $\rho$ from 
Section~\ref{sec:projectionAlignment}, since all occurrences of~$\rho$ in this paper correspond to a correlation coefficient whose evaluation formula is clear from its arguments. Finally, we define the objective function 
\begin{equation}\label{eq:refinement objective}
	c(\theta) = 1-\rho(T_{\Theta}(\phi_{1}),\phi_{2}),
\end{equation}
which vanishes for the parameters $\Theta$ that align $\phi_{1}$ with $\phi_{2}$.

To refine~$O_{est}$ and~$t_{est}$ of Section~\ref{sec:Rotations}, we simply apply the BFGS algorithm~\cite{BFGS} to the objective function~\eqref{eq:refinement objective}, with an initialization given by $O_{est}$ and~$t_{est}$.

\section{Results}\label{sec:results}

The alignment algorithm (with and without the optional refinement described in Section~\ref{sec:optimization}) was implemented in Python and is available online\footnote{\url{https://github.com/ShkolniskyLab/emalign}}, including the code that generates the figures of this section. A Matlab version of the algorithm is available as part of the ASPIRE software package~\cite{Aspire}. 

As the algorithm uses two parameters -- the downsampling $n_{ds}$ (see Section~\ref{sec:implementation}) and the number of reference projections $N$ (see Section~\ref{sec:Rotations}) -- we first examine how to appropriately set their values. Then, we examine the advantage of the refinement procedure proposed in Section~\ref{sec:optimization}. To show the benefits of our algorithm in practice, we then compare its performance to that of two other alignment algorithms -- the alignment  algorithm from the EMAN2 software package~\cite{EMAN2} (implemented in the program \texttt{e2proc3d}) and the fast rotational matching algorithm implemented in the Xmipp software package~\cite{Xmipp}. Finally, we examine the performance of the three algorithms using noisy input volumes.

We tested our algorithm on volumes from the electron microscopy data bank (EMDB)~\cite{EMDB} with different types of symmetries, whose properties are described in Table~\ref{table:datasets}. All tests were executed on a dual Intel Xeon E5-2683 CPU (32 cores in total), with 768GB of RAM running Linux. The memory required by the algorithm is of the order of the size of the input volumes. We used $15,236$  candidate rotations in Algorithm~\ref{al:Align2D} (the size of the set~$S$), generated as described in Appendix~\ref{app:candidates}. This set of candidates is roughly equally spaced in the set of rotations $SO(3)$. While it is difficult to characterize the resolution of this set in terms of the resolution of each of the Euler angles, a rough calculation suggests that the resolution in each of the Euler angles is smaller than 5~degrees. We do not use  rotations generated by a regular grid of Euler angles, as such a grid is less efficient than our grid, due to the nonuniform rotations generated by a regular grid of Euler angles. For example, discretizing each of the Euler angles to 5~degrees would result in 186,624 rotations, more than an order of magnitude larger than the number of rotations we use.

\begin{table}
	\centering
	\begin{tabular}{rlr} \hline
		EMDID & Sym & Size ($n$)\\
		\hline
		2660 &	C1  & 360\\ 
		0667 & C2  & 480\\
		0731 & C3  & 486\\
		0882 & C4  & 160\\
		21376 & C5 & 256\\
		11516 & C7 & 512\\
		21143 & C8 & 256\\
		6458 & C11 & 448\\
		30913 & D2 & 110 \\
		20016 & D3 & 384\\
		22462 & D4 & 320 \\
		9233 & D7  & 400\\
		21140 & D11&  324\\
		4179 & T  & 200 \\
		24494 & I  & 432\\
		\hline
	\end{tabular}
	\caption{Test volumes. Each volume is a three-dimensional array of size $n\times n\times n$, with $n$ specified on the third column. The symmetry of each volume is given by the second column.} 
	\label{table:datasets}
\end{table}

For each test, we generate a pair of volumes $\phi_1$ and $\phi_2$ related by a rotation matrix $O$ and a translation vector $t\in \mathbb{R}^3$. The translation is chosen at random with magnitude up to 10\% of the size of the volume. We denote the alignment parameters estimated by our algorithm by~$O_{est}$ and~$t_{est}$.   We evaluate the accuracy of our algorithm by calculating the difference between the rotations $O$ and $O_{est}$. To that end, we first note that following~\eqref{eq:sym_est_for_O}, $O_{est}$ is an estimate of $gO$ for some arbitrary $g\in G_1$, where $G_1\subseteq SO(3)$ is the symmetry group of $\phi_1$.
In order to calculate the difference between $O$ and $O_{est}$, we have to find the symmetry element~$g$. 
In our tests, the symmetry group $G_1$ is known (see Table~\ref{table:datasets}), and so we find $g$ by solving
\begin{equation}\label{eq:res:g_known}
	\argmin_{g\in G_1} \norm{O_{est} - gO}_{F},
\end{equation}
followed by defining $O_{est}' = g^TO_{est}$. 
Next, the error in the estimated rotation $O'_{est}$ is calculated using the axis-angle representation of rotations as follows.
The axis of the rotation $O$ is defined to be the unit vector $v\in \mathbb{R}^3$ that satisfies $Ov=v$, that is, $v$ is an eigenvector of $O$ corresponding to  eigenvalue~$1$. Similarly, we define the unit vector $v'\in \mathbb{R}^3$ to be the axis of the rotation $O_{est}'$.
Then, we calculate the angle between the axes of the rotations as
\begin{equation}\label{eq:e_1}
	e_1 = \cos^{-1}(v^{T} v').
\end{equation}
The angle of rotation of the matrix $O$ around its axis $v$ is given by $\theta_1 = \cos^{-1}(u \cdot Ou)$, where $u\in \mathbb{R}^3$ is a unit vector perpendicular to $v$. Similarly, we define $\theta_{2}$ to be 
the angle of rotation of the matrix $O_{est}'$ around its axis $v'$. 
The error in the rotation angle is then defined as  
\begin{equation}\label{eq:e_2}
	e_2 = |\theta_1 - \theta_2|.
\end{equation}

\begin{figure}
	\centering
	\includegraphics[width=0.5\textwidth]{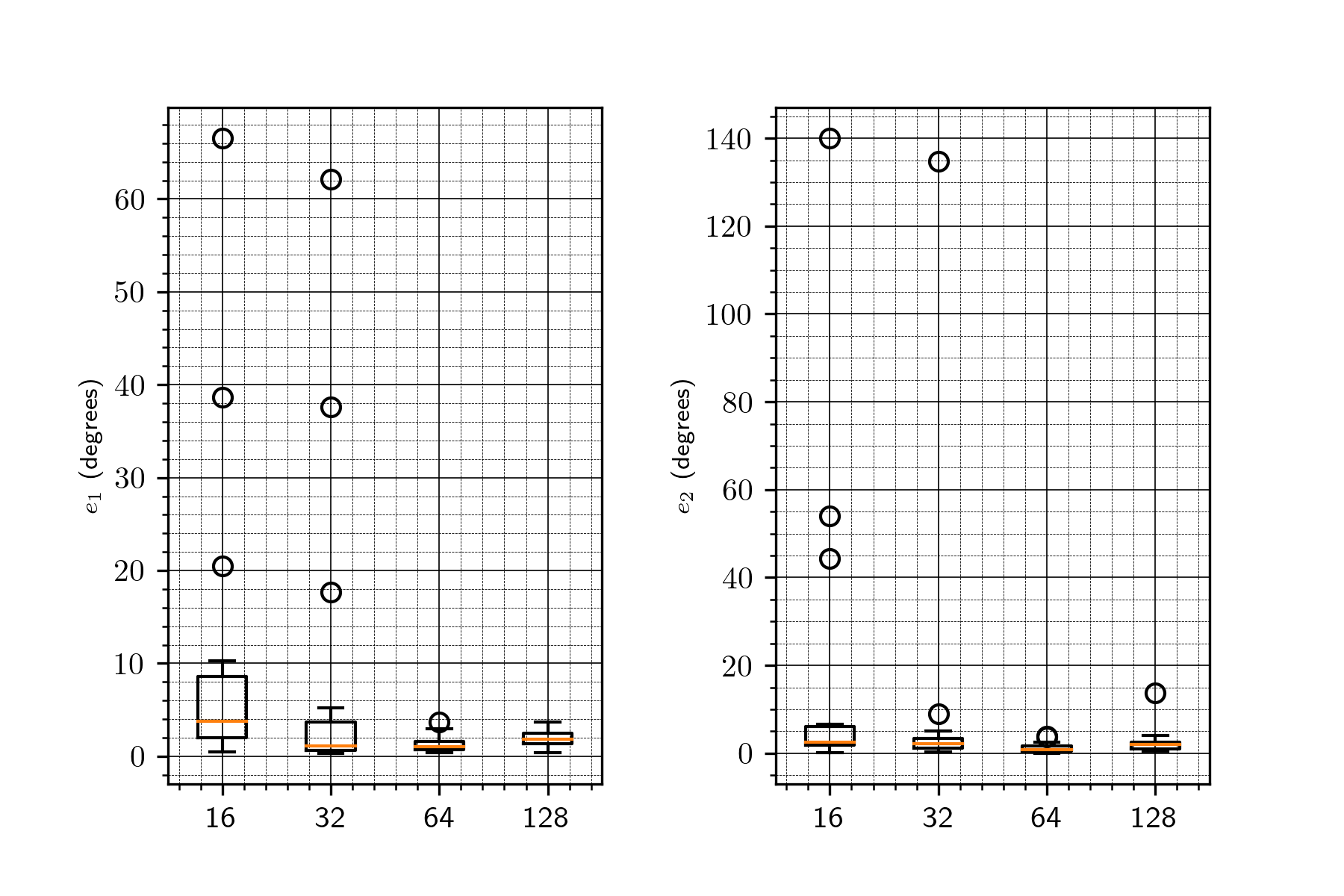}
	\caption{Downsampling parameter vs. accuracy of the algorithm. The left figure corresponds to the error $e_{1}$ in the rotation axis (see~\eqref{eq:e_1}). The right figure corresponds to the error $e_{2}$ in the rotation angle  (see~\eqref{eq:e_2}).}
		\label{fig:downsampling accuracy}
\end{figure}

\begin{figure}
	\centering
	\includegraphics[width=0.5\textwidth]{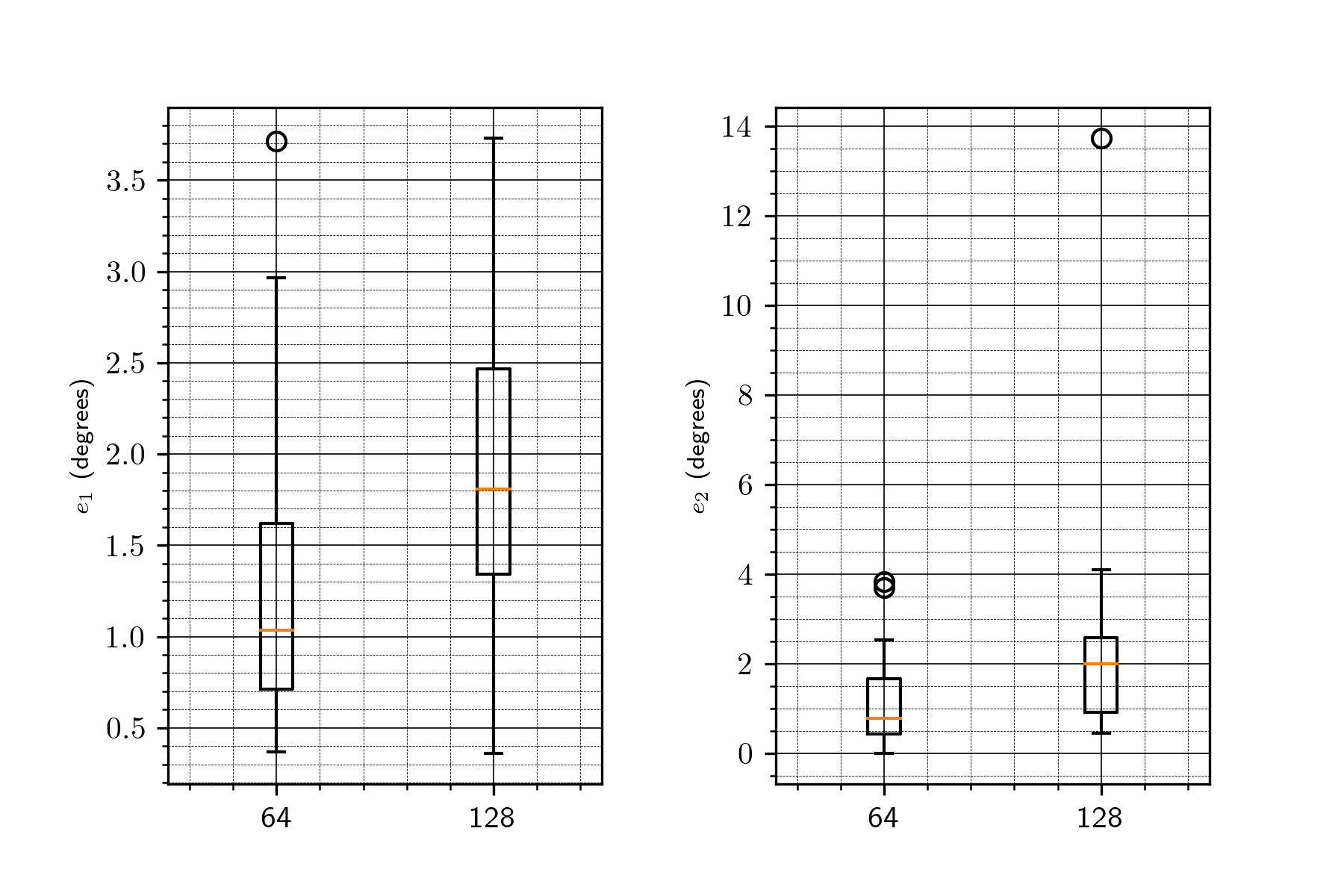}
	\caption{Downsampling parameter vs. accuracy of the algorithm, focused on 64 and 128. See Fig.~\protect\ref{fig:downsampling accuracy} for more details.}
	\label{fig:downsampling accuracy focused}
\end{figure}

\begin{figure}
	\centering
	\includegraphics[width=0.5\textwidth]{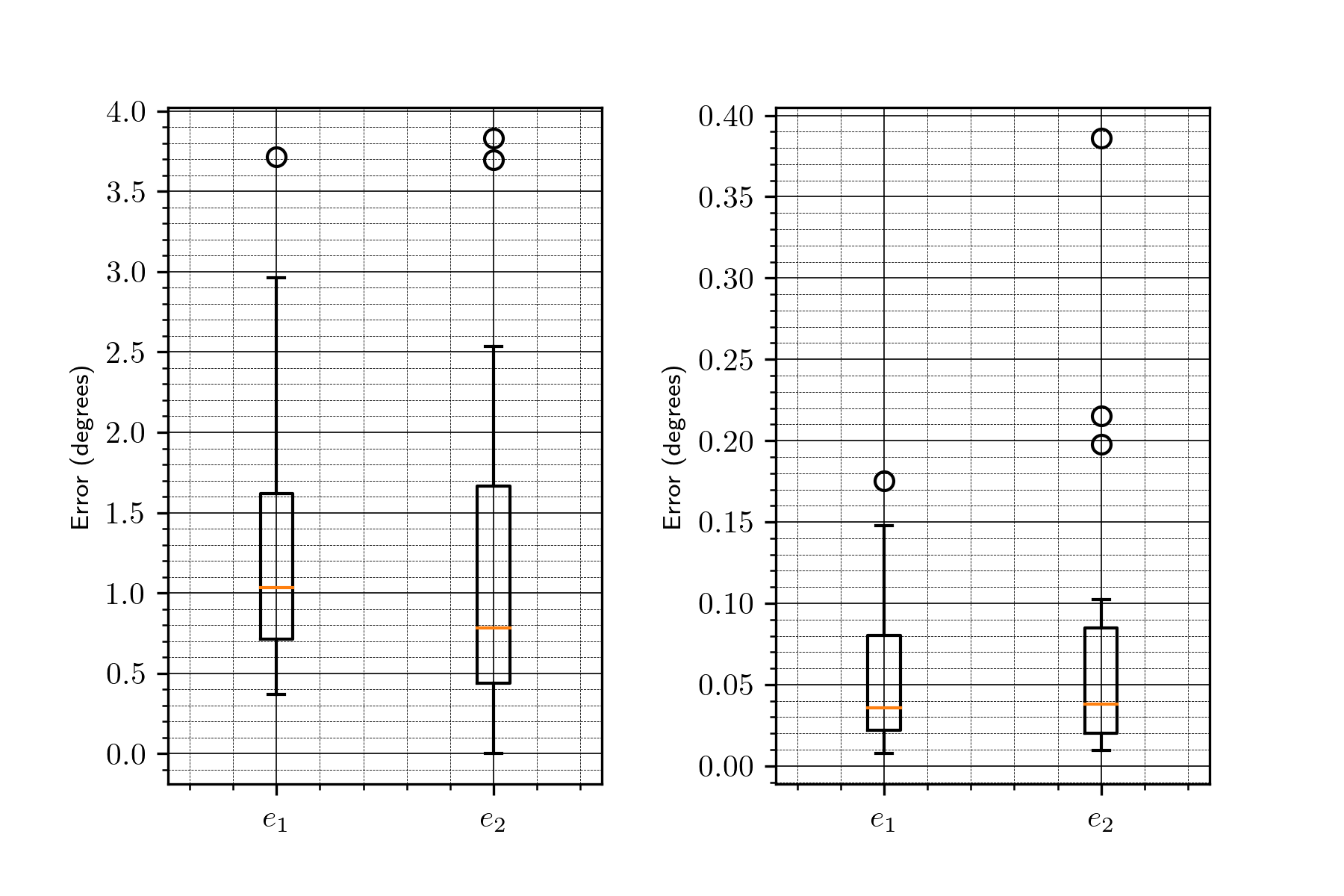}
	\caption{Error without (left figure) and with (right figure) refinement for downsampling to size $64\times 64\times 64$. The error reported in the figure is either $e_{1}$~\protect\eqref{eq:e_1} or~$e_{2}$~\eqref{eq:e_2}, as shown on the $x$-axis.}
	\label{fig:downsampling accuracy refine}
\end{figure}

\begin{figure}
	\centering
	\includegraphics[width=0.5\textwidth]{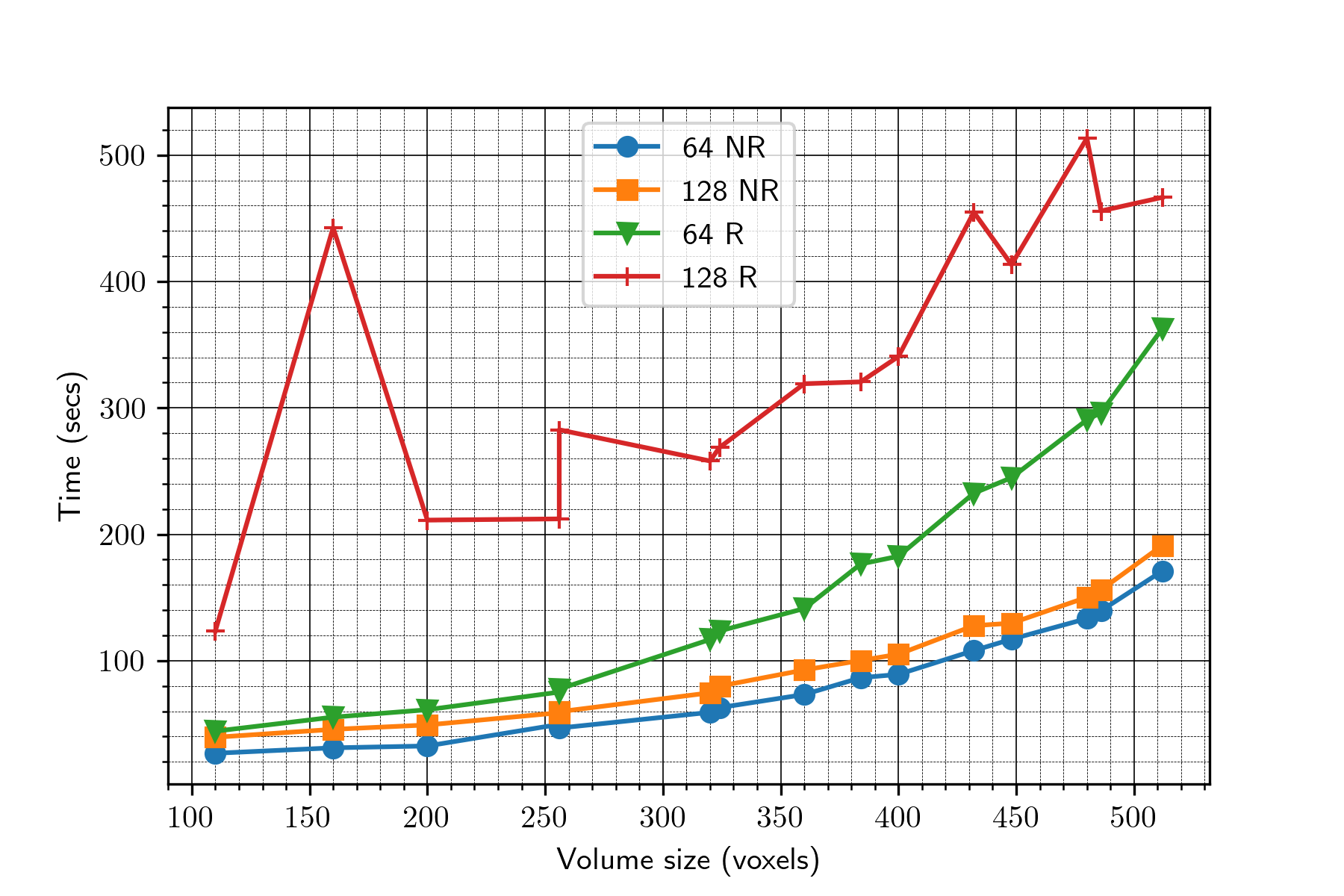}
	\caption{Timing of the alignment algorithm with downsampling to sizes 64 and 128. NR stands for ``without refinement''; R stands for ``with refinement''.}
	\label{fig:downsampling timing}
\end{figure}

We start by investigating the appropriate value for the downsampling parameter $n_{ds}$ (see Section~\ref{sec:implementation}). To that end, for each of the volumes in Table~\ref{table:datasets}, we create its rotated and shifted copy, and apply our algorithm with the downsampling parameter equal to 16, 32, 64, and 128 (namely, we actually align downsampled copies of the volumes and then apply the estimated parameters to the original volumes). The results are shown in Fig.~\ref{fig:downsampling accuracy}. For each value of downsampling, we show a bar plot that summarizes the results for all test volumes. Note that these results are without the refinement procedure of Section~\ref{sec:optimization}. To provide a more detailed information on the chosen downsampling value, we show in Fig.~\ref{fig:downsampling accuracy focused} only the results for downsampling to sizes~64 and~128. Based on these results, we use a downsampling value of~64 in all subsequent tests. In particular, this value of downsampling results in an accurate initialization of the refinement procedure of Section~\ref{sec:optimization}, as shown in Fig.~\ref{fig:downsampling accuracy refine}. As of timing, we show in Fig.~\ref{fig:downsampling timing} the timing, without and with refinement, for downsampling to sizes~64 and~128.

\begin{figure}
	\centering
	\includegraphics[width=0.5\textwidth]{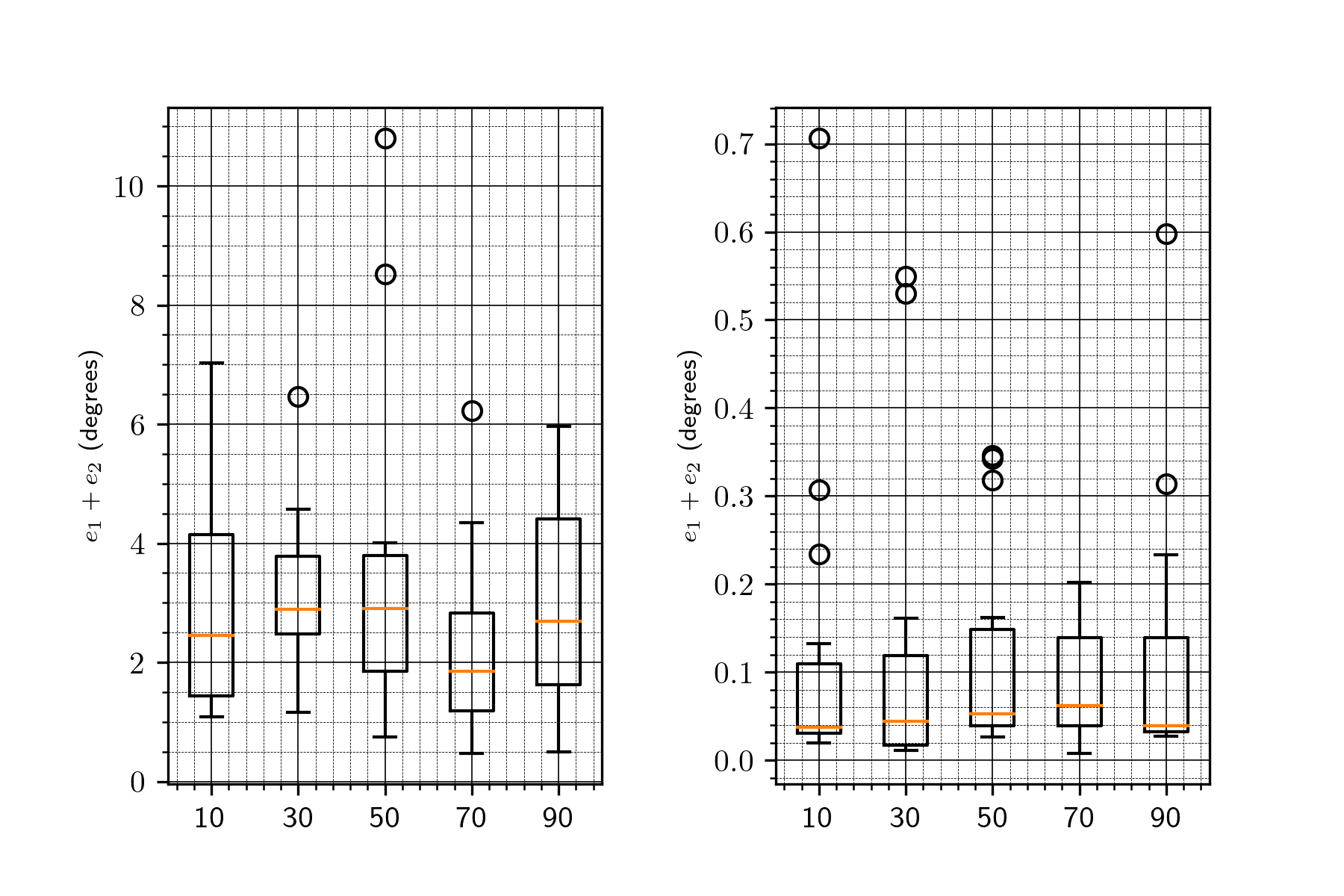}
	\caption{Error vs. the number of reference projections $N$. The left and right figures show the error without and with the refinement procedure of Section~\ref{sec:optimization}, respectively. The error reported in this figure is the sum $e_{1}+e_{2}$ given in~\protect\eqref{eq:e_1} and~\eqref{eq:e_2}.}
	\label{fig:nprojs accuracy}
\end{figure}

\begin{figure}
	\centering
	\includegraphics[width=0.5\textwidth]{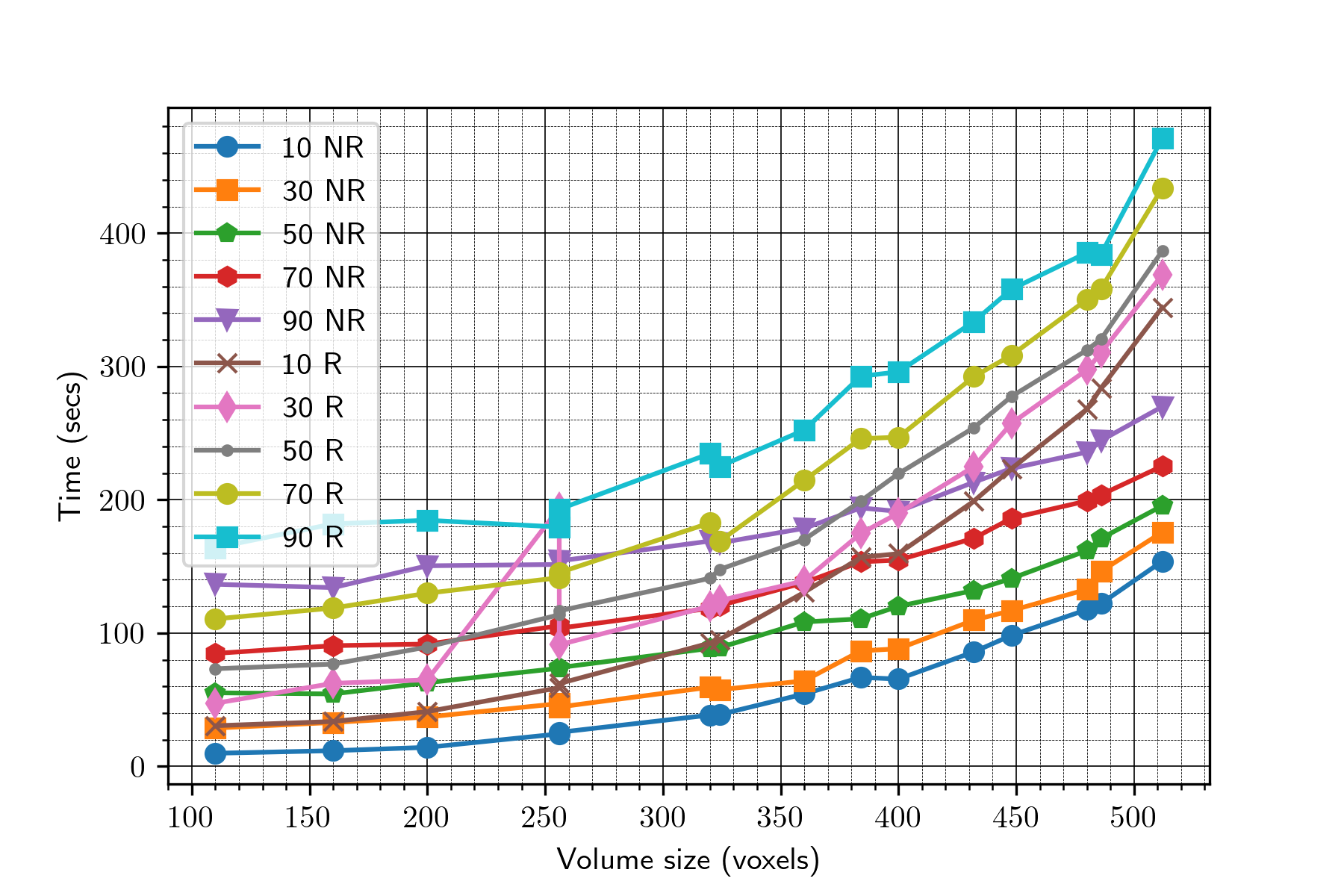}
	\caption{Time vs. the number of reference projections.}
	\label{fig:nprojs timing}
\end{figure}

Next, we wish to determine the number of reference projections $N$ to use in Algorithms~\ref{al:case_1} and~\ref{al:Align2D}. We set the downsampling parameter to $64$, and measure the estimation error for different numbers of reference projections. The results are summarized in Fig.~\ref{fig:nprojs accuracy}. We also show the timing for different numbers of reference projections, without and with refinement, in Fig.~\ref{fig:nprojs timing}. Based on these results, we choose the number of reference projections to be 30, as a good compromise between accuracy and speed.

\begin{figure}
\begin{center}
\subfloat[Clean]{\includegraphics[width=0.22\textwidth]{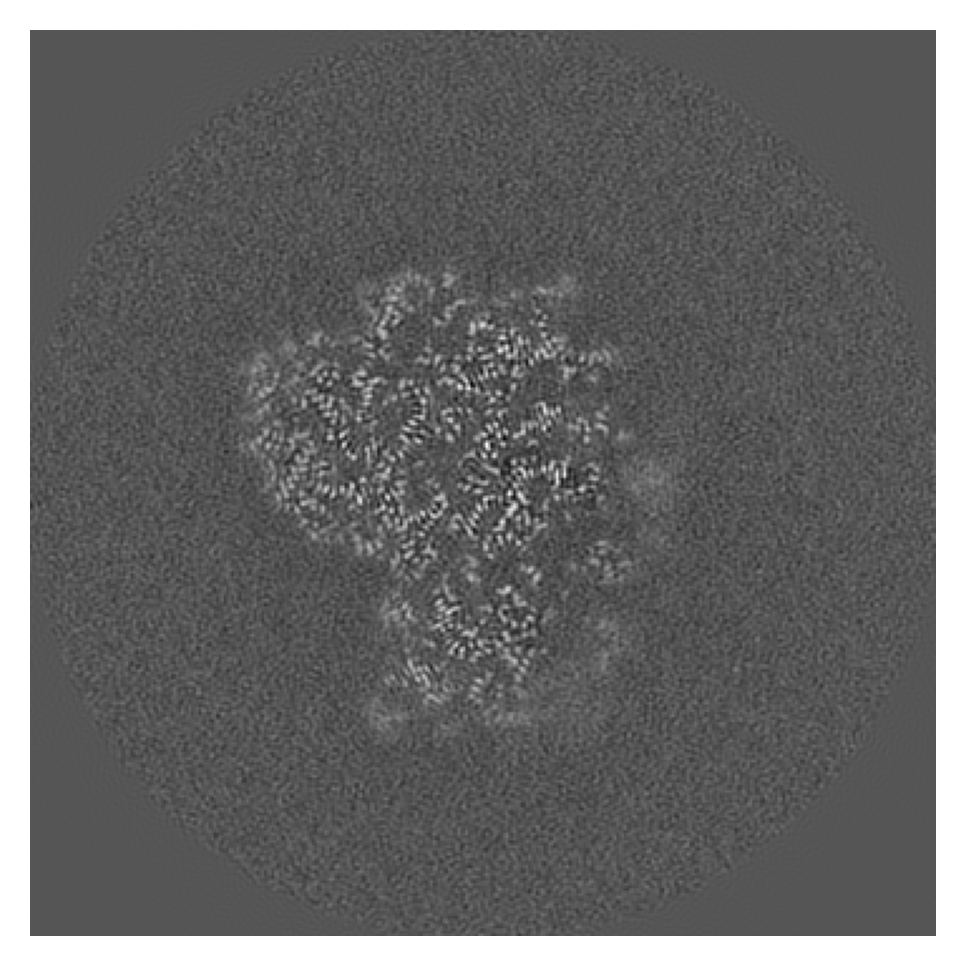}}%
\subfloat[SNR=1]{\includegraphics[width=0.22\textwidth]{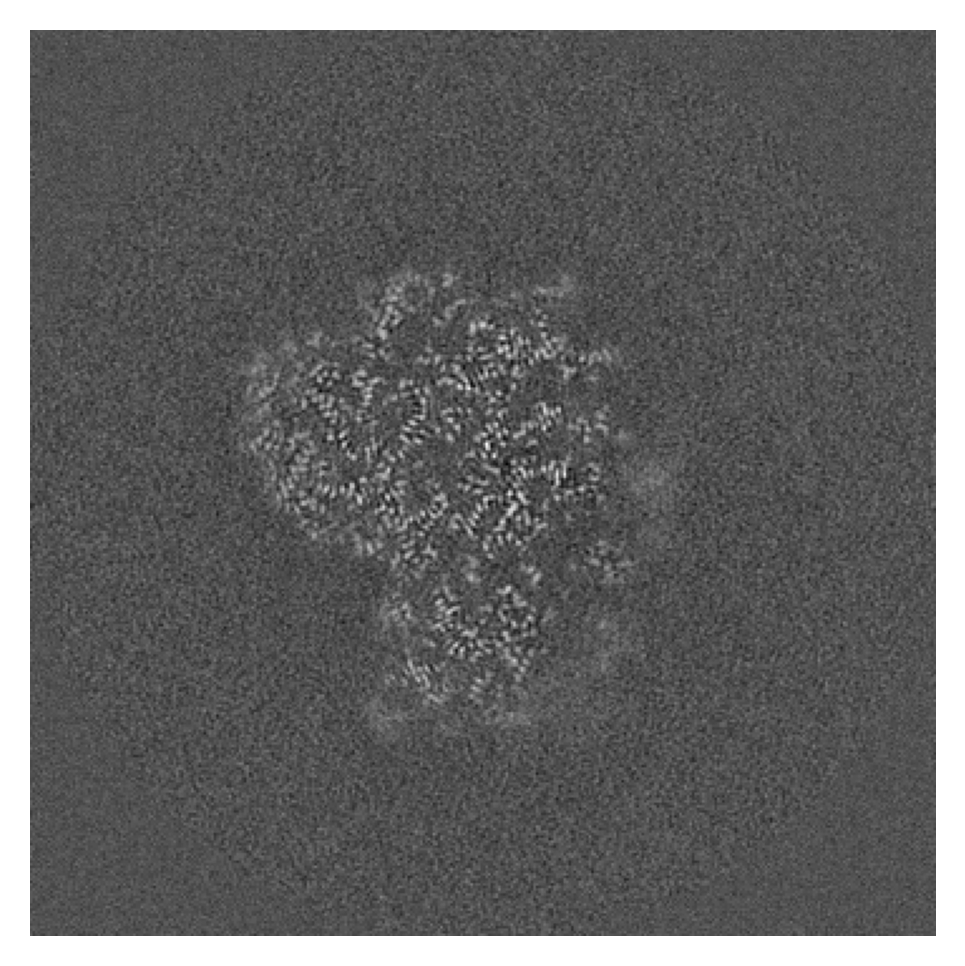}}%
\subfloat[SNR=1/2]{\includegraphics[width=0.22\textwidth]{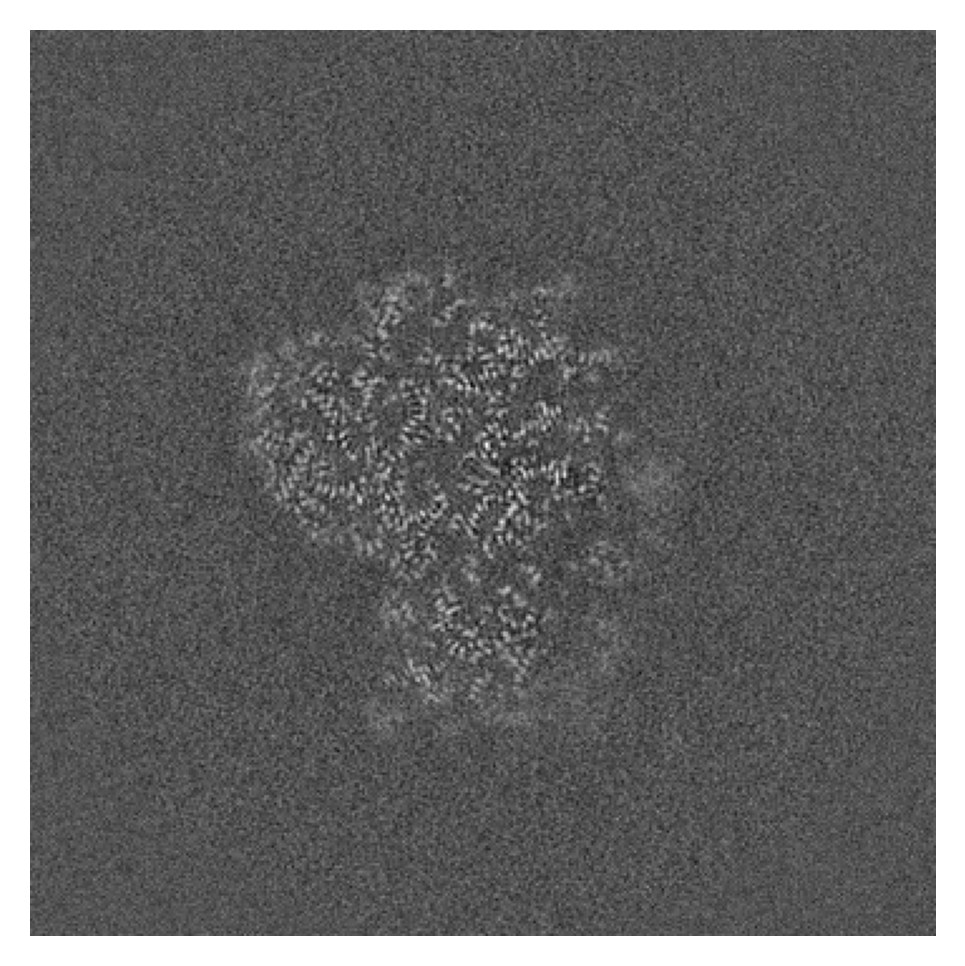}}%
\subfloat[SNR=1/8]{\includegraphics[width=0.22\textwidth]{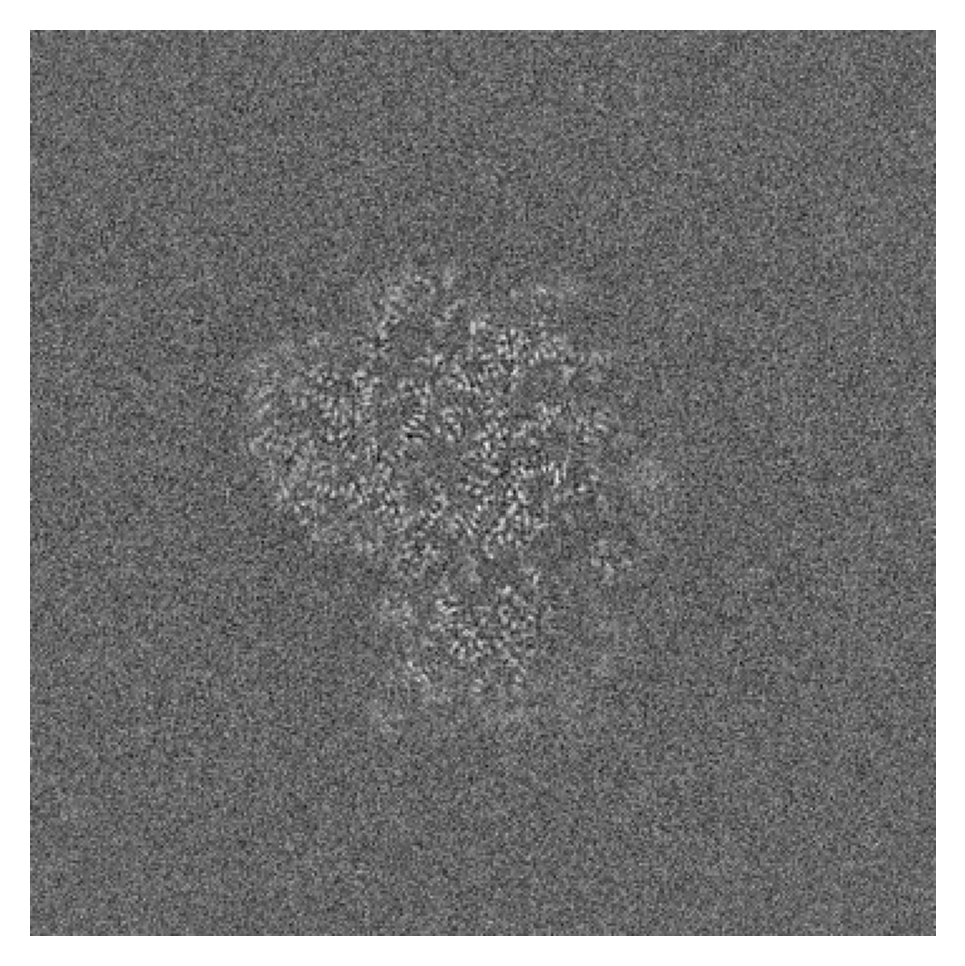}}\\
\subfloat[SNR=1/32]{\includegraphics[width=0.22\textwidth]{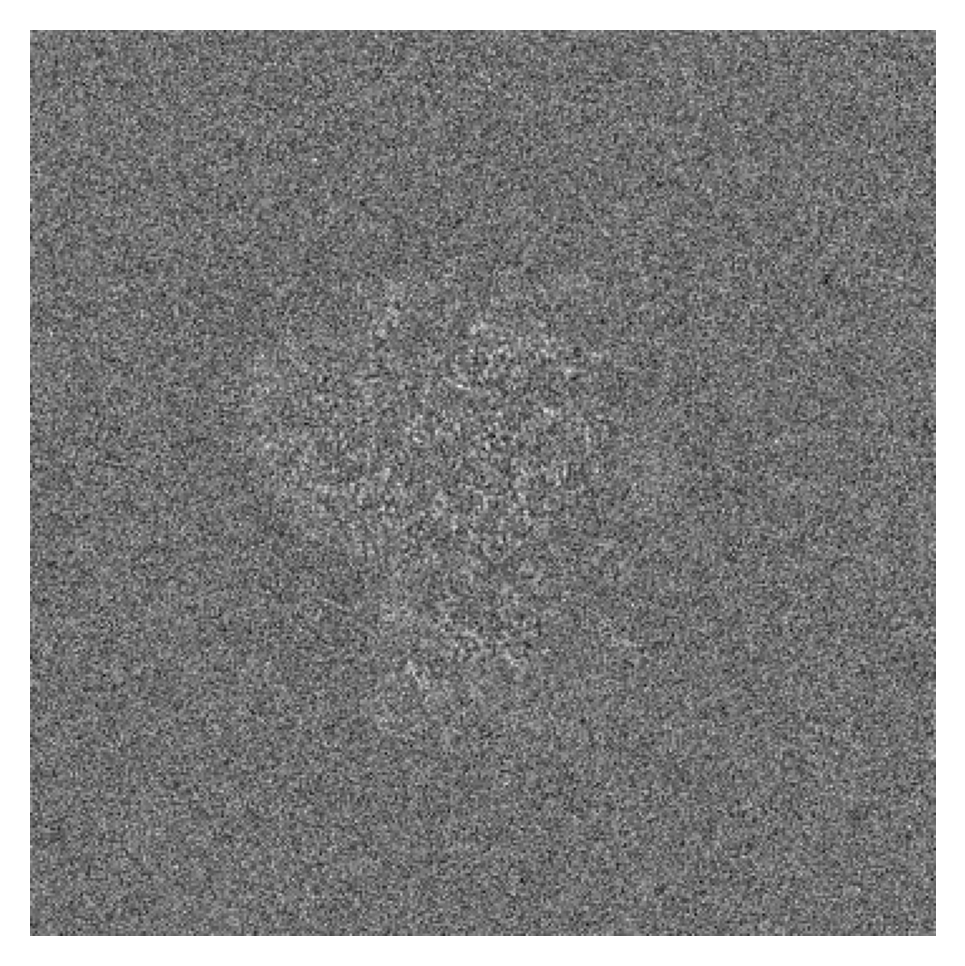}}
\subfloat[SNR=1/64]{\includegraphics[width=0.22\textwidth]{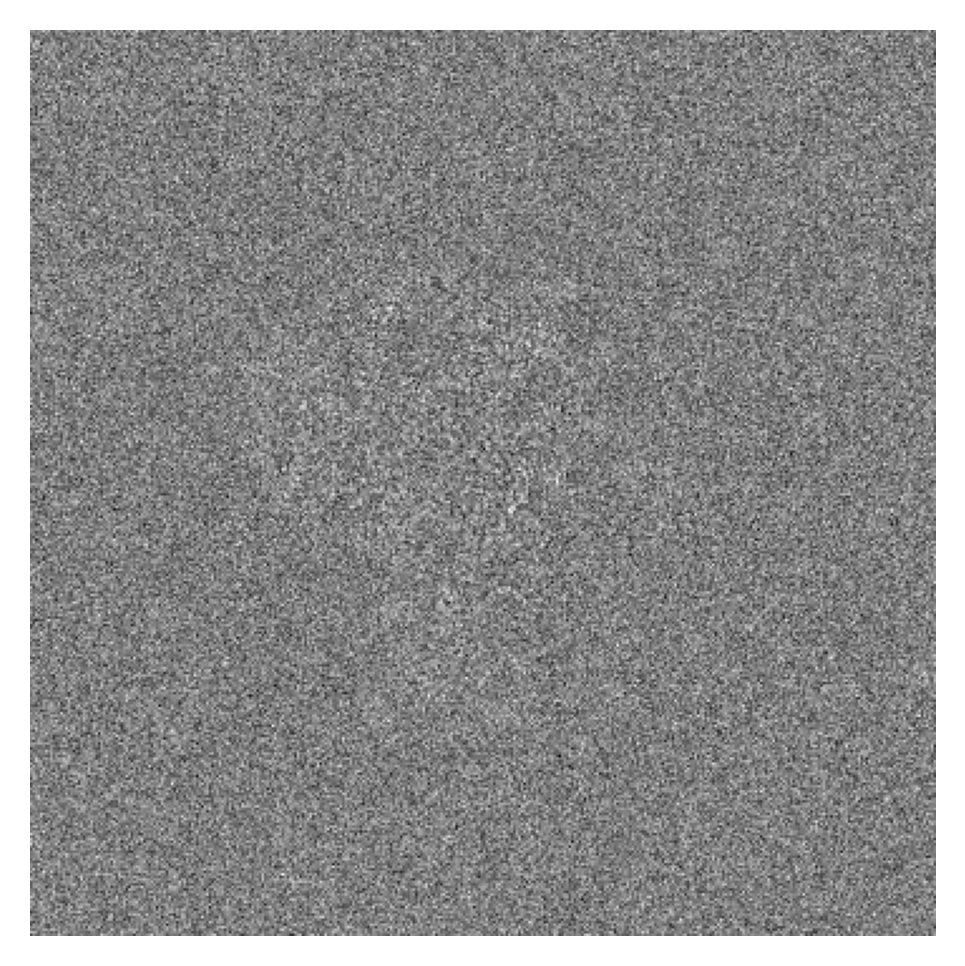}}%
\subfloat[SNR=1/128]{\includegraphics[width=0.22\textwidth]{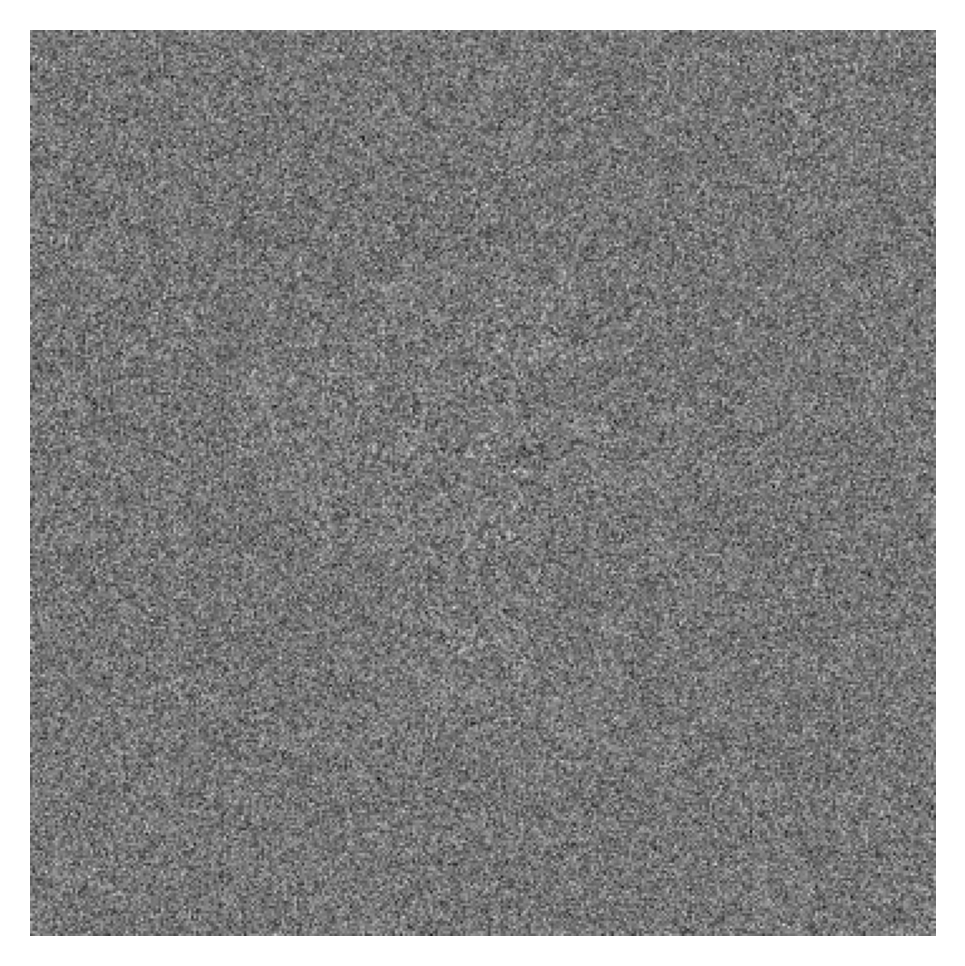}}%
\subfloat[SNR=1/256]{\includegraphics[width=0.22\textwidth]{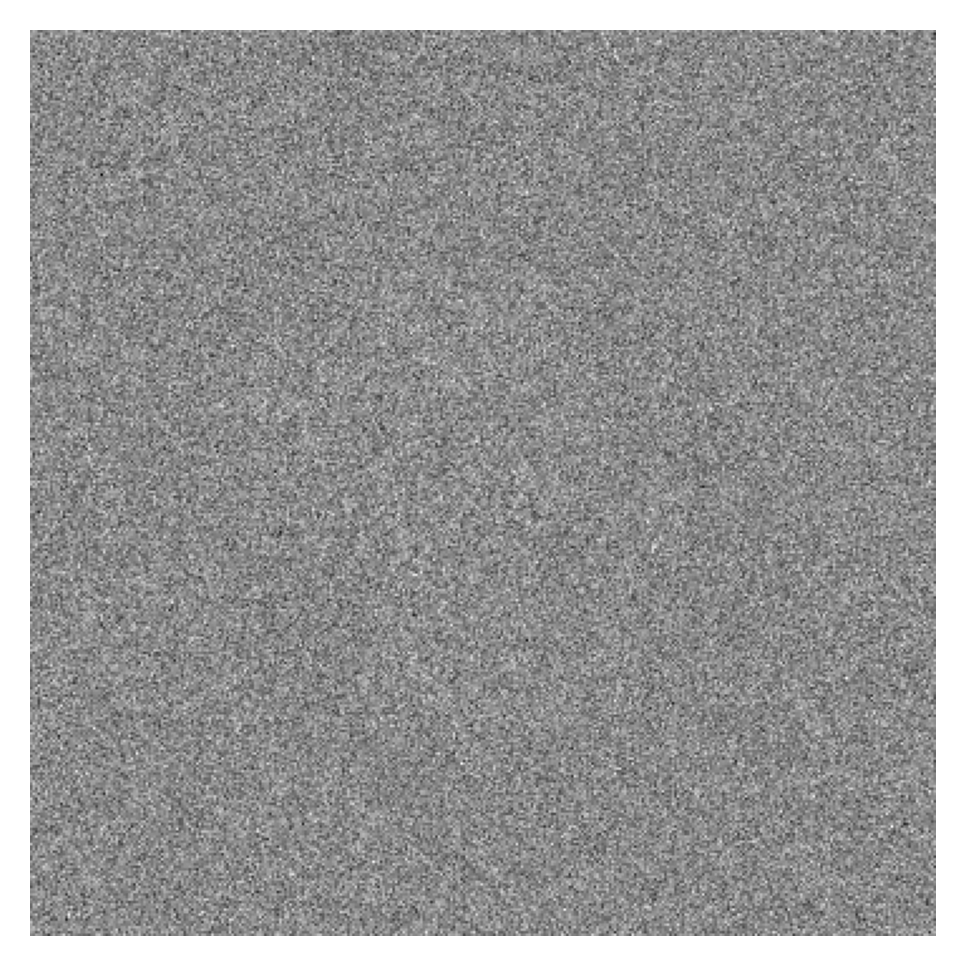}}%
\end{center}
\caption{Central slice of the noisy reference volume at different SNRs.}
\label{fig:noisy slices}
\end{figure}

Next, we compare the performance of our algorithm with that of EMAN2's and Xmipp's alignment algorithms. The accuracy and timing results are summarized in 
Tables~\ref{tbl:comparisons accuracy} and~\ref{tbl:comparisons timing}, respectively. Finally, we demonstrate the performance of the different algorithms for noisy input volumes. To that end, we  use as a reference volume EMD~2660~\cite{10.7554/eLife.03080} from EMDB (of size $360\times 360\times 360$ voxels), and create its rotated and translated copy. We add to the reference volume and its rotate/translated copy additive Gaussian noise with SNR ranging from~1 to~1/256. A central slice from the noisy reference volume at different levels of SNR is shown in Fig.~\ref{fig:noisy slices}. The accuracy results of all algorithms for the various SNRs are shown in Table~\ref{tbl:snr accuracy}. The timings of the different algorithms are shown in Table~\ref{tbl:snr timing}.

\begin{table}
\begin{center}
\begin{tabular}{l r r r r r r}
	\toprule
	Sym & EMDID & EMalign(NR) & EMalign(R) &  EMAN &  Xmipp \\
	\midrule
	C1 &  2660 &       2.802 &      0.094 & 0.094 &  5.557 \\
	C2 &   667 &       3.747 &      0.223 & 0.121 &  7.181 \\
	C3 &   731 &       8.664 &      5.131 & 0.135 & 48.058 \\
	C4 &   882 &       1.952 &      0.041 & 0.408 &  0.317 \\
	C5 & 21376 &       2.949 &      0.358 & 0.507 & 15.445 \\
	C7 & 11516 &       3.961 &      0.397 & 0.153 &  5.745 \\
	C8 & 21143 &       1.502 &      0.536 & 0.455 &  2.778 \\
	C11 &  6458 &       2.825 &      0.116 & 0.046 &  0.314 \\
	D2 & 30913 &       6.273 &      0.035 & 0.425 &  0.141 \\
	D3 & 20016 &       3.499 &      0.075 & 0.033 &  1.564 \\
	D4 & 22462 &       6.016 &      0.126 & 0.095 &  0.251 \\
	D7 &  9233 &       4.034 &      0.063 & 0.029 &  5.866 \\
	D11 & 21140 &       3.183 &      0.042 & 0.247 &  0.127 \\
	T &  4179 &       1.324 &      0.556 & 0.348 &  6.246 \\
	I & 24494 &       3.268 &      0.030 & 0.114 &  0.028 \\
	\midrule
	mean &       &       3.733 &      0.522 & 0.214 &  6.641 \\
	std &       &       1.945 &      1.288 & 0.168 & 12.206 \\
	\bottomrule
\end{tabular}
\end{center}
\caption{Accuracy comparison with EMAN2 and Xmipp. The errors reported in this table are the sum $e_{1}+e_{2}$ given in~\protect\eqref{eq:e_1} and~\eqref{eq:e_2}. Errors are given in degrees. For EMalign, (NR) corresponds to ``without refinement'' and (R) to ``with refinement''. The two bottom rows show the mean and standard deviation of the error (in degrees) over all experiments.}
\label{tbl:comparisons accuracy}
\end{table}

\begin{table}
\begin{center}
\begin{tabular}{l r r r r r r}
	\toprule
	Sym & EMDID & size & EMalign(NR) & EMalign(R) & EMAN & Xmipp \\
	\midrule
	C1 &  2660 &  360 &          49 &        130 &  172 &  2106 \\
	C2 &   667 &  480 &          80 &        235 &  354 &  5812 \\
	C3 &   731 &  486 &          85 &        173 &  351 &  5582 \\
	C4 &   882 &  160 &          18 &         55 &   66 &    91 \\
	C5 & 21376 &  256 &          24 &         58 &  155 &   529 \\
	C7 & 11516 &  512 &          78 &        216 &  425 &  6854 \\
	C8 & 21143 &  256 &          33 &         55 &  120 &   698 \\
	C11 &  6458 &  448 &          54 &        151 &  276 &  3854 \\
	D2 & 30913 &  110 &          16 &         34 &   55 &    37 \\
	D3 & 20016 &  384 &          41 &        105 &  197 &  2214 \\
	D4 & 22462 &  320 &          29 &         87 &  201 &  1095 \\
	D7 &  9233 &  400 &          40 &        124 &  171 &  2970 \\
	D11 & 21140 &  324 &          35 &         79 &  197 &  1175 \\
	T &  4179 &  200 &          21 &         59 &   80 &   246 \\
	I & 24494 &  432 &          54 &        158 &  281 &  3313 \\
	\bottomrule
\end{tabular}
\end{center}
\caption{Timing comparison with EMAN2 and Xmipp (in seconds). For EMalign, (NR) corresponds to ``without refinement'' and (R) to ``with refinement''. The column ``size'' is the side length of the input volumes.}
\label{tbl:comparisons timing}
\end{table}

\begin{table}
\begin{center}
\begin{tabular}{l r r r r r r r r}
	\toprule
	SNR & EMalign(NR) & EMalign(R) &  EMAN & Xmipp \\
	\midrule
	clean &       4.066 &      0.143 & 0.072 & 0.968 \\
	1 &       3.715 &      0.145 & 0.072 & 0.898 \\
	1/2 &       1.827 &      0.150 & 0.072 & 0.851 \\
	1/8 &       5.733 &      0.291 & 0.072 & 0.728 \\
	1/32 &       5.014 &      3.318 & 0.095 & 0.811 \\
	1/64 &       4.283 &      0.598 & 0.105 & 1.124 \\
	1/128 &       2.727 &      0.691 & 0.202 & 1.177 \\
	1/256 &       4.449 &     25.089 & 0.124 & 1.598 \\
	1/512 &      92.569 &     97.549 & 0.288 & 1.662 \\
	\bottomrule
\end{tabular}
\end{center}
\caption{Accuracy comparison for noisy input volumes at different SNRs. See Table~\ref{tbl:comparisons accuracy} for more details.}
\label{tbl:snr accuracy}
\end{table}

\begin{table}
	\begin{center}
\begin{tabular}{l r r r r r r r r}
	\toprule
	SNR & EMalign(NR) & EMalign(R) & EMAN & Xmipp \\
	\midrule
	clean &          33 &        102 &  181 &  2097 \\
	1 &          35 &        122 &  176 &  2091 \\
	1/2 &          39 &         97 &  175 &  2177 \\
	1/8 &          33 &        105 &  170 &  2121 \\
	1/32 &          33 &        100 &  155 &  1991 \\
	1/64 &          38 &         97 &  157 &  2125 \\
	1/128 &          39 &        113 &  179 &  2297 \\
	1/256 &          36 &        101 &  163 &  2092 \\
	1/512 &          39 &         92 &  167 &  2365 \\
	\bottomrule
\end{tabular}

	\end{center}
	\caption{Timing comparison for noisy input volumes at different SNRs. All timings are given in seconds.}
	\label{tbl:snr timing}
\end{table}

\section{Discussion and conclusions}\label{sec:conclusions}
In this paper, we proposed a fully automatic method for aligning three-dimensional  volumes with respect to rotation, translation, and reflection. While the parameters of the algorithm can be tuned whenever needed, we showed that the default parameters work very well for a wide range of volumes of various symmetries. We also developed an auxiliary algorithm which finds the orientation of a volume giving rise to a given projection image (Section~\ref{sec:projectionAlignment}). This algorithm may serve as a fast and highly accurate substitute to projection matching. 

The core difference between our approach and other existing approaches is that our approach is based on commons line between projection images generated from the volumes. 
The advantage of this approach is that inspecting each candidate rotation is very fast, as it is based on one-dimensional operations on the common lines ($O(n)$ operations for volumes of size $n\times n\times n$). We also note that our cost function~\eqref{eq:cost function} for identifying the optimal alignment is different than in other algorithms. While the typical cost function used by alignment algorithms is the correlation between the volumes, our cost function is the average correlation of the common lines between projection images of the volumes. These two cost functions are not equivalent, and while in our experiments we have not identified a scenario where one cost function is superior over the other, having tools that are based on different principles may prove beneficial in the future.

From the comparison of our algorithm with the alignment algorithms in EMAN2 and Xmipp, we conclude  that our algorithm can be used in one of two modes.
If we are interested in fast alignment with good accuracy (average error of 
1.9~degrees of the rotation axis, and average error of 1.86~degrees of the in-plane rotation angle, with standard deviations of 1.25~degrees and 1.3~degrees, respectively), we can use our algorithm without the refinement procedure of Section~\ref{sec:optimization}. This is appropriate, for example, for visualization, as such an initial alignment is sufficient as an input for high resolution optimization-based alignment algorithms, such as the one in Chimera~\cite{chimera}. In such a case, our algorithm is more than~3~times faster than EMAN2's algorithm (even though our algorithm is implemented entirely in Python), and almost~40~times faster than Xmipp's algorithm. If we are interested in very low alignment errors, the refinement procedure of Section~\ref{sec:optimization} brings the average errors down to 0.25~degrees for the rotation axis and	0.28~degrees for the in-plane rotation angle (with standard deviations of 0.66~degrees and 0.63~degrees, respectively). In such a case, our algorithm is 80\% faster than EMAN2's and 15 times faster than Xmipp's.

\section*{Acknowledgments}
This research was supported by the European Research Council (ERC) under the European Union's Horizon 2020 research and innovation programme (grant
agreement 723991 - CRYOMATH) and by the NIH/NIGMS Award R01GM136780-01.

\appendix

\section{Common lines}\label{app:common lines}
In this section, we review the Fourier projection slice theorem and its induced common line property. Given a volume~$\phi$ and a rotation matrix~$R$, the projection image of $\phi$ corresponding to orientation $R$ is given by~\eqref{eq:projImage}. We identify~$R^{(3)}$ (the third column  of $R$) as the viewing direction of~$\phi$ (see~\cite{singer2011ViewingAngle}). The first two columns $R^{(1)}$ and $R^{(2)}$ of $R$ form an orthonormal basis for a plane in $\mathbb{R}^3$ which is perpendicular to the viewing direction $R^{(3)}$. Therefore, if $R_i$ and $R_j$ are two rotations with the same viewing direction ($R_i^{(3)}=R_j^{(3)}$), then the two projection images $P_{i}$ and $P_{j}$ generated according to~\eqref{eq:projImage} look the same up to some in-plane rotation.

The Fourier projection slice theorem relates the two-dimensional Fourier transform of a projection image~$P$ with the three-dimensional Fourier transform of~$\phi$. 
Let
\begin{equation*}
	\hat{P}(\omega_x,\omega_y) = \iint_{\mathbb{R}^2} P(x,y)e^{-\imath(x\omega_x+y\omega_y)}dxdy
\end{equation*}
be the two-dimensional Fourier transform of $P(x,y)$, and let
\begin{equation*}
	\hat{\phi}(\omega_x,\omega_y,\omega_z) = \iiint_{\mathbb{R}^3} \phi(x,y,z)e^{-\imath(x\omega_x+y\omega_y+z\omega_z)}dxdydz 
\end{equation*}
be the three-dimensional Fourier transform of $\phi(x,y,z)$.
The Fourier projection slice theorem~\cite{shkolnisky2012viewing} states that
\begin{equation}\label{proj-sliceThem}
	\hat{P}(\omega_x,\omega_y) =  \hat{\phi}(\omega_x R^{(1)}+\omega_y R^{(2)}),
\end{equation}
where $P$ is defined in \eqref{eq:projImage}.
Equation \eqref{proj-sliceThem} states that the two-dimensional Fourier transform of each projection image $P$ is equal to a planar slice of the three-dimensional Fourier transform of $\phi$. Moreover, it states that this planar slice is the plane $\omega_x R^{(1)}+\omega_y R^{(2)}$. The Fourier projection slice theorem~\eqref{proj-sliceThem} holds, up to discretization errors, also for discrete volumes and their sampled projection images.

From \eqref{proj-sliceThem}, we get that any two Fourier transformed projection images $\hat{P}_{i}$ and $\hat{P}_{j}$ with different viewing directions ($R_i^{(3)} \neq R_j^{(3)})$ are equal to two different planar slices from~$\hat{\phi}$. Since there exists a line that is common to both planar slices, the two Fourier transformed images share a common line. We refer to that line as the common line between $P_{i}$ and $P_{j}$. We denote the angle that this line makes with the local $x$-axis of the (Fourier transformed) images $\hat{P}_{i}$ and $\hat{P}_{j}$ by $\alpha_{ij}$ and $\alpha_{ji}$, respectively. 
Mathematically, the common line property is expressed as~\cite{shkolnisky2012viewing} 
\begin{equation*}
	\hat{P}_{i}(\xi \cos \alpha_{ij}, \xi \sin \alpha_{ij}) = \hat{P}_{j}(\xi \cos \alpha_{ji}, \xi \sin \alpha_{ji}),\quad  \xi\in\mathbb{R},
\end{equation*}   
implying that the samples of the Fourier transformed images along the common line are equal.

To find an expression for the angles $\alpha_{ij}$ and $\alpha_{ji}$, we consider the unit vector 
\begin{equation*}
	q_{ij} = \frac{ R_i^{(3)} \times R_j^{(3)} }{||R_i^{(3)} \times R_j^{(3)}||},
\end{equation*}
where $\times$ is the cross product between vectors. Define the unit vectors 
\begin{equation*}
		c_{ij} = (\cos \alpha_{ij}, \sin \alpha_{ij},0 )^T, \quad c_{ji} = (\cos \alpha_{ji}, \sin \alpha_{ji},0 )^T.
\end{equation*}
It can be shown~\cite{shkolnisky2012viewing} that these vectors satisfy the equation
\begin{equation*}
	R_ic_{ij} = q_{ij} = R_jc_{ji},
\end{equation*}
which implies that $c_{ij}$ and $c_{ji}$ can be computed as
\begin{equation*}
	c_{ij} = R_{i}^{T}q_{ij},\quad c_{ji} = R_{i}^{T}q_{ij},
\end{equation*}
from which $\alpha_{ij}$ and $\alpha_{ji}$ can be easily extracted.

\section{Constructing the set $S$}\label{app:candidates}
We generate the set of candidate rotations $S$  by using the Euler angles representation for rotations.
Let~$L$ be a positive integer, and let $\tau,\theta,\varphi$ be Euler angles. We construct $S$ by sampling the Euler angles in equally spaced increments as follows. First, we sample $\tau \in \{0,\ldots,\frac{\pi}{2}\}$ at $\lfloor \frac{L}{4}\rfloor$ points. Then, for each $\tau$, we sample $\theta \in \{0,\ldots, \pi\}$ at  $\lfloor \frac{L}{2}\sin(\tau)\rfloor$ points. Finally, for each pair $(\tau,\theta)$, we sample  $\varphi \in \{0,\ldots, 2\pi\}$ at $\lfloor \frac{L}{2}\sin(\tau)\sin(\theta)\rfloor$ points. For each $(\tau,\theta,\varphi)$ on this grid, we compute a corresponding rotation matrix~$R$ by 
\begin{equation*}
    R = R_z(\tau)R_y(\theta)R_x(\varphi),
\end{equation*}
where
\begin{align*}
R_z(\tau) &= \begin{pmatrix}
\cos\tau & -\sin\tau & 0\\
\sin\tau & \cos\tau & 0\\
0 & 0 & 1\\ 
\end{pmatrix},\\  
R_y(\theta) &= \begin{pmatrix}
\cos\theta & 0 & \sin\theta  \\
0 & 1 & 0\\
-\sin\theta & 0 & \cos\theta \\
\end{pmatrix},\\  
R_x(\varphi)&= \begin{pmatrix}
1 & 0 & 0\\
0 & \cos\varphi & -\sin\varphi \\
0 & \sin\varphi & \cos\varphi \\
\end{pmatrix}.
\end{align*}

\section{Translation estimation}\label{app:ShiftEst3D}
For completeness, we review the well-known phase correlation procedure for translation estimation~\cite{phase_corr}. Consider two volumes $\phi_1$ and $\phi_2$ shifted relative to one another, that is 
\begin{equation}
    \phi_2(r) = \phi_1(r-t), \quad r=(x,y,z)^T\in \mathbb{R}^3,
\end{equation}
where $t=(\Delta_x,\Delta_y,\Delta_z)^T\in \mathbb{R}^3$. Our goal is to estimate $t$. 

First, by the Fourier shift property, the Fourier transforms of $\phi_{1}$ and $\phi_{2}$ satisfy 
\begin{equation}\label{eq:3D_shift_property}
    \hat{\phi}_2(\omega_x,\omega_y,\omega_z) =  \hat{\phi}_1(\omega_x,\omega_y,\omega_z)e^{-\imath(\omega_x\Delta_x+\omega_y\Delta_y+\omega_z\Delta_z)}.
\end{equation}
From~\eqref{eq:3D_shift_property} we get~\cite{506761}
\begin{equation}\label{eq:cross-power}
	\begin{aligned}
    \hat{\rho}(\omega_x, \omega_y, \omega_z) &= \frac{\hat{\phi}_1  {\hat{\phi}_2}^*}{\lvert \hat{\phi}_1  {\hat{\phi}_2}^* \rvert} = \frac{\hat{\phi}_1 \hat{\phi}_1^* e^{\imath(\omega_x\Delta_x+\omega_y\Delta_y+\omega_z\Delta_z)}}{\lvert \hat{\phi}_1 \hat{\phi}_1^* \rvert} \\
    &=  e^{\imath(\omega_x\Delta_x+\omega_y\Delta_y+\omega_z\Delta_z)},
    \end{aligned}
\end{equation}
since $\lvert e^{\imath(\omega_x\Delta_x+\omega_y\Delta_y+\omega_z\Delta_z)}\rvert = 1$. 
Then, since the inverse Fourier transform of a complex exponential is a Dirac delta, we have
\begin{equation}
    \rho(x,y,z) = \delta(x+\Delta_x,y+\Delta_y,z+\Delta_z).
\end{equation}
Therefore, $t=(\Delta_x,\Delta_y,\Delta_z)^T$ is given by 
\begin{equation}\label{eq:t_3D}
    (\Delta_x,\Delta_y,\Delta_z) = -\argmax_{(x,y,z)} \rho(x,y,z).
\end{equation}

While this appendix is formulated in the continuous domain, the same holds if we replace $\phi_{1}$ and $\phi_{2}$ by their discrete versions sampled on a regular grid, and replace the Fourier transform by the discrete Fourier transform.

\bibliographystyle{plain}
\bibliography{main}

\end{document}